\documentclass[conf]{new-aiaa}
\usepackage[utf8]{inputenc}

\usepackage{graphicx}
\usepackage{amsmath}
\usepackage[version=4]{mhchem}
\usepackage{siunitx}
\usepackage{longtable,tabularx}
\usepackage{subcaption} %
\usepackage{booktabs}
\usepackage{float}
\usepackage{verbatim}
\usepackage{natbib} %

\title{DSMC Study on Unsteadiness of Supersonic Flow Over a Cylinder}
\author{Mert Senkardesler\footnote{Graduate Student, Department of Aerospace Engineering, AIAA Student Member.}, Irmak T. Karpuzcu\footnote{Graduate Student, Department of Aerospace Engineering, AIAA Student Member.} and Deborah A. Levin\footnote{Professor, Department of Aerospace Engineering, AIAA Fellow}}
\affil{University of Illinois, Urbana-Champaign,IL 61801,USA}

\date{December 17, 2025}

\begin{document}

\maketitle
\begingroup\renewcommand{\thefootnote}{}\footnotetext{Presented at the AIAA SciTech 2026 Forum, 12-16 January 2026, Orlando, FL. Published by the American Institute of Aeronautics and Astronautics, Inc. Cite as: M. Senkardesler, I. T. Karpuzcu, and D. A. Levin, ``DSMC Study on Unsteadiness of Supersonic Flow Over a Cylinder,'' AIAA Paper 2026-2147, AIAA SciTech 2026 Forum, Orlando, FL, January 2026, \url{https://doi.org/10.2514/6.2026-2147}. Copyright \textcopyright{} 2026 by the American Institute of Aeronautics and Astronautics, Inc. All rights reserved. This is the authors' submitted version, posted in accordance with the AIAA self-archiving and posting policy.}\addtocounter{footnote}{-1}\endgroup

\begin{abstract}
  Unsteady flow over a \SI{30}{\micro\metre} diameter cylinder at Mach 4 is investigated using Direct Simulation Monte Carlo across Reynolds numbers from 270 to 6750, spanning slip-flow to near-continuum regimes. Power Spectral Density analysis at discrete probe locations and Dynamic Mode Decomposition of flowfield snapshots independently characterize instabilities in both the stagnation and wake regions.
  In the stagnation region, a bow shock pulsation mode at 10-15~MHz frequency is identified across all cases despite a 24-fold variation in number density.
  In the wake, two distinct instabilities are observed. The canonical vortex shedding mode at Strouhal number \( St \approx 0.2 \) dominates at lower Reynolds numbers. A second instability at \( St \approx 0.5 \), consistent with previously reported slipline-driven oscillations in supersonic wakes, emerges at higher Reynolds numbers and progressively strengthens, becoming the dominant mode.
  In the studied Reynolds range, the wake transitions from no clear unsteadiness, to canonical vortex shedding at \( St \approx 0.20 \), and reaching a regime where the \( St \approx 0.5 \) mode becomes dominant, while also having a distinct independent bow shock pulsation possibly governed by its standoff distance.
\end{abstract}

\newpage
\section{Introduction}
\begin{figure}[H]
  \centering
  \includegraphics[width=0.85\linewidth]{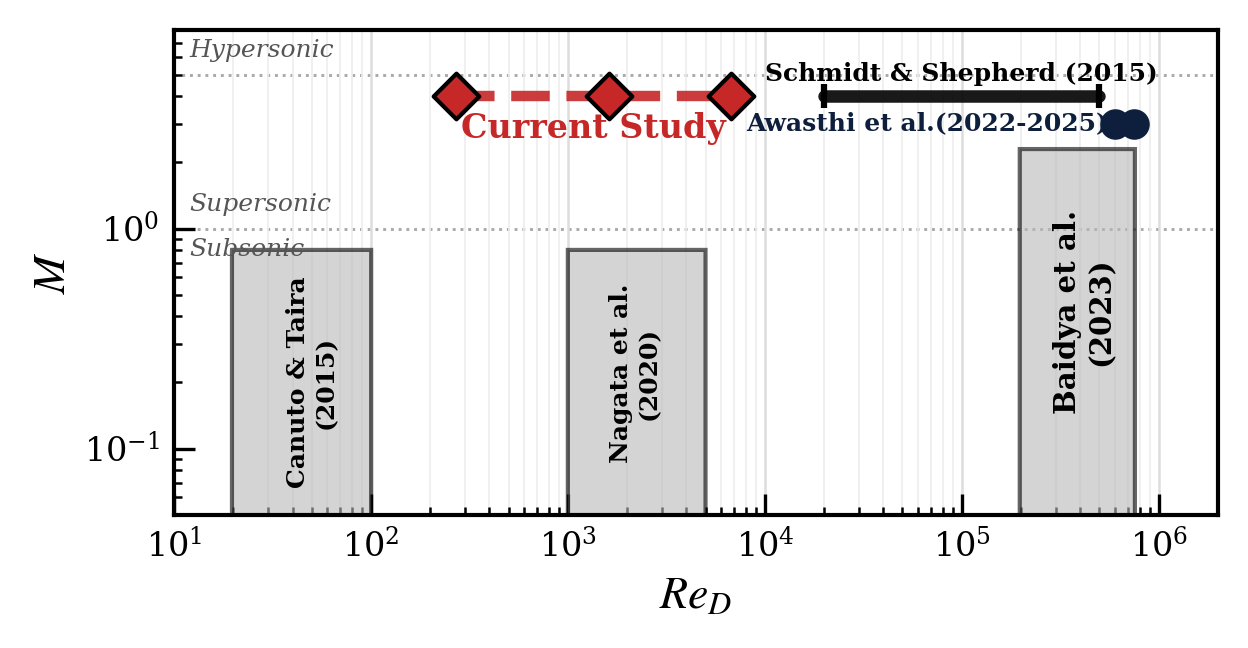}
  \caption{Map of Mach number versus Reynolds number for previous experimental and numerical studies on flow over a circular cylinder~\citep{canuto_two-dimensional_2015, nagataExperimentalInvestigationCompressible2020, schmidtOscillationsCylinderWakes2015, awasthiSupersonicCylinderWake2022, baidya_investigation_2023,awasthi2025coherent}. The regime investigated in the current study is highlighted, occupying a sparsely explored region of high Mach number and low-to-moderate Reynolds number.}
  \label{fig:mach_reynolds_regime}
\end{figure}

The behavior of flow around a circular cylinder at high Mach numbers (e.g., Mach 4) but low Reynolds numbers is deceptively rich in detail and behaves fundamentally differently from its well-charted counterparts in the subsonic or much higher Reynolds regimes. The current study initially originated from a simple but probing question: what happens to an atmospheric particulate caught in near-hypersonic flow? Such a condition places us in a high-Mach, low-Reynolds, rarefied regime where standard intuition begins to fail. A defining feature of this regime is the combination of compressibility and rarefaction effects. To understand the context, we must view the flow regime in the Mach-Reynolds space, as shown in Figure \ref{fig:mach_reynolds_regime}. This map reveals territory between the well-understood domains, highlighting the gap this study aims to fill.

The current study occupies the high-Mach, low-Reynolds, high-Knudsen number regime. Literature on this regime is sparse, which is not surprising, since intuition dictates that compressibility and rarefaction would suppress any unsteadiness, rendering this regime uninteresting. The initial observations that motivated the current study involved persistent unsteadiness in the stagnation region and bow shock, which was suspected to be a coupled response to wake instability. However, the wake was quiet, and the stagnation unsteadiness persisted even in truncated domains with no wake at all. To check for wake contribution, the study was therefore devised to span a range of Reynolds numbers where wake unsteadiness would be expected. In doing so, it was found that the wake would exhibit multiple distinct unsteady modes in this reynolds range, a finding that agrees with recent developments in the literature.

For flow over a cylinder, the canonical Kármán vortex shedding ($St \approx 0.2$) is the dominant instability, ubiquitous in subsonic flows. However, as compressibility effects are introduced with increasing Mach numbers, wake unsteadiness is suppressed, a phenomenon showcased by Canuto and Taira~\citep{canuto_two-dimensional_2015} for subsonic flow ($M < 0.5$) at low Reynolds numbers ($Re = 20$--$100$). Their results show that increasing the Mach number unambiguously raises the critical Reynolds number ($Re_c$), stabilizing the primary wake instability even in the simplified regime of shock-free, laminar flow.
At much higher Reynolds numbers, the picture is similar. Experimental Schlieren visualizations by Baidya et al.~\citep{baidya_investigation_2023} show that while subsonic wakes remain wide with interacting shear layers, high Mach numbers force these layers to converge toward the centerline. This results in a physically confined wake which is an order of magnitude narrower than its subsonic equivalent, fundamentally inhibiting the interaction required for the canonical Kármán vortex street.

One might therefore conclude that supersonic wakes are universally stable. However, experimental work by Schmidt and Shepherd~\citep{schmidtOscillationsCylinderWakes2015} challenged this assumption. In their study of a cylinder at Mach 4 for Reynolds numbers from $2 \times 10^4$ to $5 \times 10^5$, they identified a dominant wake oscillation distinct from the Kármán vortex street, observing a Strouhal number of $St \approx 0.42$. They hypothesized that this unsteadiness arises from interactions between the shear layers and the wake neck, where the neck width, rather than the cylinder diameter, becomes the governing length scale. This mechanism is fundamentally different from the viscosity-driven shear layer roll-up of subsonic flows, establishing the existence of a new, unsteady mode in supersonic wakes.

The mechanism proposed by Schmidt and Shepherd requires the formation of a supersonic wake characterized by sliplines converging to a neck. One can then reason that even in subsonic flows but at higher Mach values, local supersonic regions and wake necks can develop, which should lead to similar instabilities. A review of the literature reveals a relevant study by Nagata et al.~\citep{nagataExperimentalInvestigationCompressible2020} on compressible flow over a cylinder at Reynolds numbers between 1000 and 5000 and Mach numbers from 0.2 to 0.5. In their PSD analysis of the wake, they observed not only the canonical vortex shedding peak at \(St \approx 0.2\) but also a distinct secondary peak near \(St \approx 0.4\), particularly at higher Mach numbers (e.g., \(Re = 4000\), \(M = 0.5\)). The authors interpreted this secondary peak as a harmonic artifact of the primary shedding frequency. However, in light of the distinct high-frequency instability mechanism now identified in compressible wakes, this interpretation warrants reconsideration. It is plausible that this mode has been observed in other studies but overlooked or misidentified as a harmonic.

Later studies by Awasthi et al.~\citep{awasthiSupersonicCylinderWake2022, awasthi2025coherent} at Mach 3--4 conclusively characterized this regime, identifying a bimodal instability where the canonical Kármán vortex shedding ($St \approx 0.2$) coexists with a dominant high-frequency oscillation ($St \approx 0.42$). Using Spectral Proper Orthogonal Decomposition (SPOD), they demonstrated that this high-frequency component is not a harmonic but a distinct aeroacoustic resonance driven by a feedback loop between the wake neck and the subsonic recirculation region, confirming the hypothesis of Schmidt and Shepherd~\citep{schmidtOscillationsCylinderWakes2015}, where they also show that whilst the hydrodynamic Kármán mode decays rapidly due to compressibility, the aeroacoustic mode persists far downstream, establishing it as the primary mechanism of wake unsteadiness.

These studies encapsulate the transition from vortex-driven to shock-driven unsteadiness, yet the regime of low-to-moderate Reynolds numbers ($Re_D \sim \mathcal{O}(10^2)-\mathcal{O}(10^4)$) at high Mach numbers remains a blind spot in the literature. Historically, the field has been bifurcated: low-Reynolds number studies are confined to subsonic flows (e.g., Nagata et al.), while high-Mach number investigations target the high-Reynolds regimes dictated by experimental facilities (e.g., Schmidt and Shepherd, Awasthi et al., where $Re_D \gtrsim 10^5$). The present study bridges this divide. By investigating the Reynolds range of $270 \le Re_D \le 6750$ at Mach 4, we operate in the same Reynolds regime as Nagata et al. but isolate the effects of Reynolds number within the compressible wake, allowing us to observe the development of the aeroacoustic instability and its coexistence with canonical vortex shedding. We employ the Direct Simulation Monte Carlo (DSMC) method to resolve these dynamics around a \SI{30}{\micro\metre} diameter cylinder, a configuration directly relevant to micro-scale atmospheric particulates interacting with hypersonic vehicles. The flowfield analysis employs both local Power Spectral Density (PSD) measurements at discrete probe locations and spatially-resolved Dynamic Mode Decomposition (DMD) of DSMC flowfield snapshots. In the stagnation region, both methods identify a distinct bow shock pulsation mode at approximately 10--15~MHz. Crucially, this unsteadiness appears to be intrinsic to the shock dynamics and independent of the wake; it persists at nearly the same frequency across all examined cases despite a 24-fold variation in number density.
In the wake region, PSD and DMD analyses reveal two distinct instabilities. The canonical vortex shedding mode (\( St \approx 0.2 \)) dominates at lower Reynolds numbers. At higher \( Re_D \), a second instability at \( St \approx 0.5 \) emerges and progressively strengthens. At \( Re_D = 6750 \), this mode dominates the transverse velocity fluctuations while the shedding mode persists in the streamwise component. The results capture the Reynolds number range over which the wake transitions from no coherent instability, through canonical vortex shedding, to a regime where the \( St \approx 0.5 \) mode becomes dominant.

\section{Simulation Setup and Flow Conditions}
This study models the unsteady flow characteristics of a \SI{30}{\micro\metre} diameter cylinder in Mach 4 flow using the Direct Simulation Monte Carlo (DSMC) method~\cite{birdMolecularGasDynamics1994}. The simulations were performed using the SMILE-DSMC solver~\cite{ivanovStatisticalSimulationMonte2012}. The methodology naturally captures stochastic thermal fluctuations, velocity slip, and temperature jump effects. Three number-density cases, ranging from \( 2.0 \times 10^{24} \,  \) to \( 5.0 \times 10^{25} \, \), are analyzed, corresponding to atmospheric conditions from sea level to 22~km altitude. For a \SI{30}{\micro\metre} diameter particulate these conditions yield cylinder-based Reynolds numbers (\( Re_D \)) of 270 to 6750 and Knudsen numbers (\( Kn_D \)) from 0.0178 to \( 7.124 \times 10^{-4} \), spanning the rarefied slip-flow to near-continuum regimes. The computational domain measures \( 130 \times 50 \, \si{\micro\metre} \), with simulations conducted over \( 2 \times 10^6 \) timesteps for a time step of \( 2.5 \times 10^{-12} \, \si{\second} \). Table~\ref{tab:flow-geometry-cases} summarizes the flow and geometry parameters.
\begin{table}[H]
  \centering
  \sisetup{
    exponent-product = \times,
    table-number-alignment = center,
    round-mode = figures,
    round-precision = 4
  }
  \caption{Flow and geometry properties for three cases at different number-densities}
  \label{tab:flow-geometry-cases}
  \renewcommand{\arraystretch}{0.75} %
  \begin{tabular}{
      l
      S[table-format=2.1]
      S[table-format=2.1]
      S[table-format=2.1]
    }
    \toprule
     & {Case 1}    & {Case 2}    & {Case 3}    \\
    \midrule
    Number density (\si{\per\cubic\metre})
     & {2.0e24}    & {1.2e25}    & {5.0e25}    \\
    Reynolds number (free-stream) (\si{\per\metre})
     & {9.00e6}    & {5.40e7}    & {2.25e8}    \\
    Mean free path (free-stream) (\si{\metre})
     & {5.3429e-7} & {8.9049e-8} & {2.1372e-8} \\
    Knudsen number
     & {0.0178}    & {0.0029}    & {7.1240e-4} \\
    \(Re_D\) (cylinder)
     & {270}       & {1620}      & {6750}      \\
    \bottomrule
  \end{tabular}
  \par
  \vspace{2pt}
  \footnotesize
  \noindent For all cases, temperature = \SI{220.0}{\kelvin}, Mach number = 4.0, cylinder diameter = \SI{30.0}{\micro\metre}, velocity = \SI{1200}{\metre\per\second}.
\end{table}

The Direct Simulation Monte Carlo (DSMC) method models gas flows by tracking the motion and collisions of representative particles, deriving macroscopic properties such as velocity, pressure, and temperature through statistical averaging within computational cells. This study employs the SMILE DSMC solver, using molecular nitrogen as the working fluid, modeled with translational, rotational, and vibrational temperatures, but excluding chemical reactions. The Variable Hard Sphere collision model is utilized to simulate intermolecular interactions, enabling accurate representation of the near-hypersonic flow around a 30 µm diameter cylinder at Mach 4 across varying number-density conditions.

\begin{figure}[H]
  \centering
  \begin{subfigure}[b]{\linewidth}
    \centering
    \includegraphics[width=0.8\linewidth]{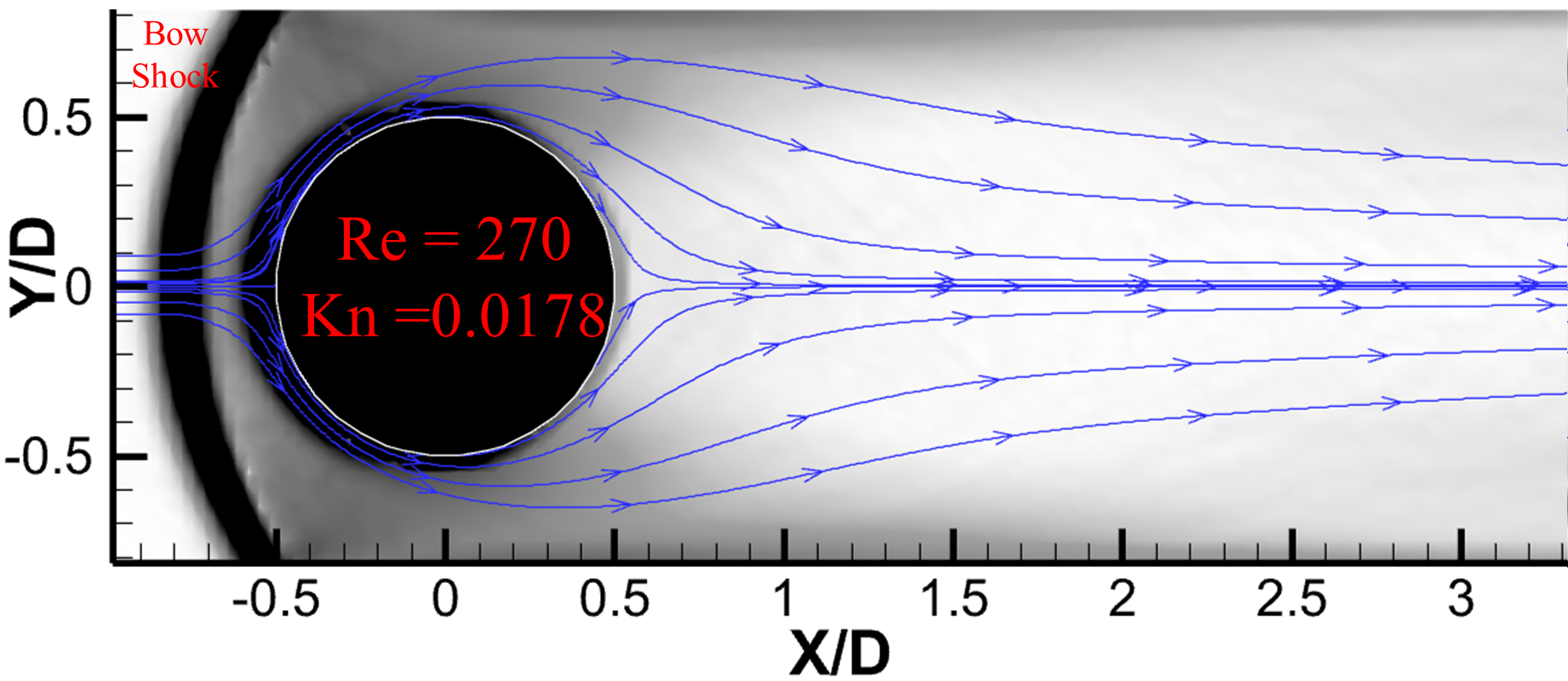}
    \caption{ N={2e24} case 1}
    \label{fig:n02}
  \end{subfigure}
  \vspace{0em} %
  \begin{subfigure}[b]{\linewidth}
    \centering
    \includegraphics[width=0.8\linewidth]{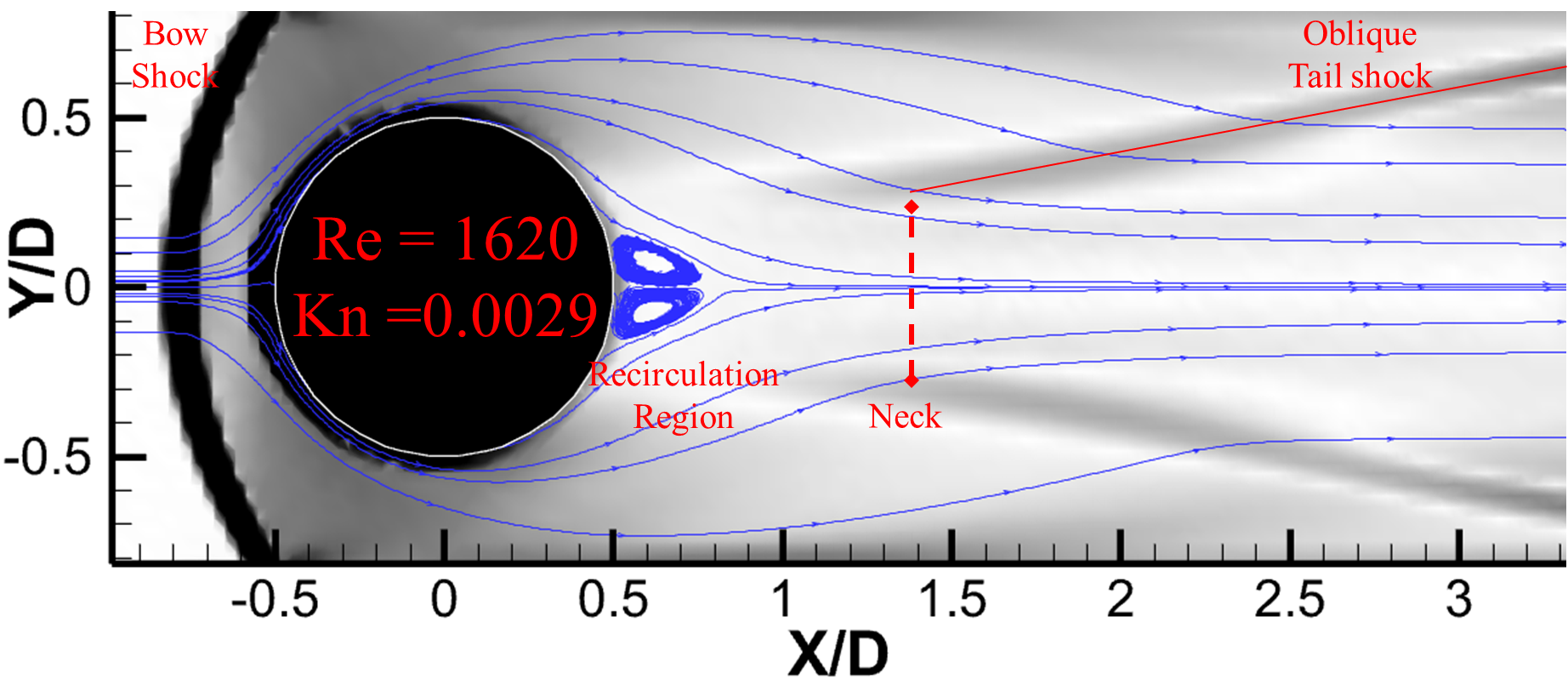}
    \caption{  N={1.2e25} case 2}
    \label{fig:n12}
  \end{subfigure}
  \vspace{0em}
  \begin{subfigure}[b]{\linewidth}
    \centering
    \includegraphics[width=0.8\linewidth]{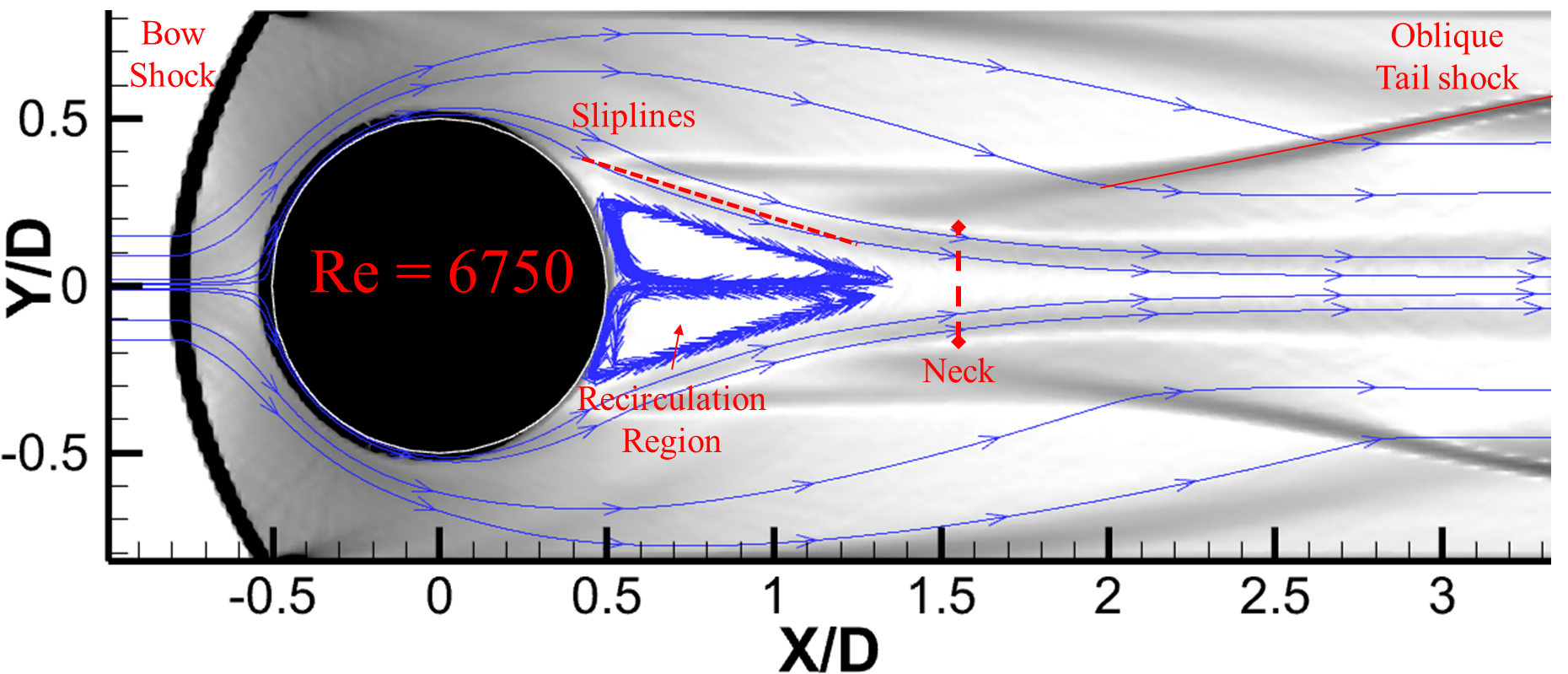}
    \caption{  N={5.0e25} case 4}
    \label{fig:n5}
  \end{subfigure}
  \caption{ Numerical schlieren illustrating Cases 1-3}
  \label{fig:schlieren-cases}
\end{figure}

Numerical schlieren images are shown in Figure~\ref{fig:schlieren-cases} to provide visual representations of density gradients around the cylinder across the three number-density cases. Figure~\ref{fig:n5}, corresponding to Case 3, shows a recirculation region with separation at the cylinder’s shoulder, sliplines converging at the neck, and an oblique tail shock, driving wake oscillations in near-continuum flows, consistent with Baidya et al.~(\citeyear{baidya_investigation_2023}). In Case 2, Figure~\ref{fig:n12} displays a prominent bow shock and recirculation, but less defined sliplines and recompression waves, i.e., it represents a more rarefied regime with weaker instabilities. In Case 1, Figure~\ref{fig:n02} reveals only a bow shock, as continuum breakdown and velocity slip, which greatly weakens boundary layer gradients, negate the recirculation and its associated instabilities.
It should be noted that a flow in the slip-flow regime does not inherently preclude recirculation in the wake. Recirculation arises when the flow cannot conform to the constraints imposed by the immersed body. In this study, velocity slip caused by insufficient molecular collisions relaxes these constraints, suppressing recirculation in Case 1. However, this is not a universal rule for flows in the slip regime. For example, \cite{zhong_comparison_2008} observed recirculation in the wake of a 10-degree cone with a sharply perpendicular base at Mach 25 and comparable Knudsen numbers. In their case, however, the recirculation originated from the Prandtl-Meyer expansion angle limit at high Mach numbers rather than boundary layer gradients, as seen in sharp shoulder geometries compared to the gradual curvature of a cylinder.

\section{Local PSD Analysis of Unsteadiness}
To quantify the unsteady behavior in the flowfield, Power Spectral Density (PSD) analysis of velocity fluctuations is performed at probe locations spanning from the stagnation region upstream of the cylinder to the far wake downstream. Instantaneous flow variables are recorded over 2 million timesteps with a timestep of \( 2.5 \times 10^{-12} \, \mathrm{s} \), and their frequency content is analyzed in the statistically steady state. This section examines two distinct regions: the bow shock ahead of the cylinder, where the DSMC method well resolves the shock layer, and the wake region, where instabilities arise from slipline dynamics and shock interactions.

\subsection{Bow Shock Dynamics}
The unsteadiness of the single detached bow shock ahead of the cylinder in hypersonic flow (\( M = 4 \)) is a flow feature that has not been extensively studied and is difficult to capture, particularly at small collisional length scales relevant to high density flows. To capture the shock's internal dynamics, probes were placed along the stagnation streamline, spanning the bow shock region ahead of the cylinder.

\begin{figure}[H]
  \centering
  \begin{subfigure}[b]{0.53\linewidth}
    \centering
    \includegraphics[width=\linewidth]{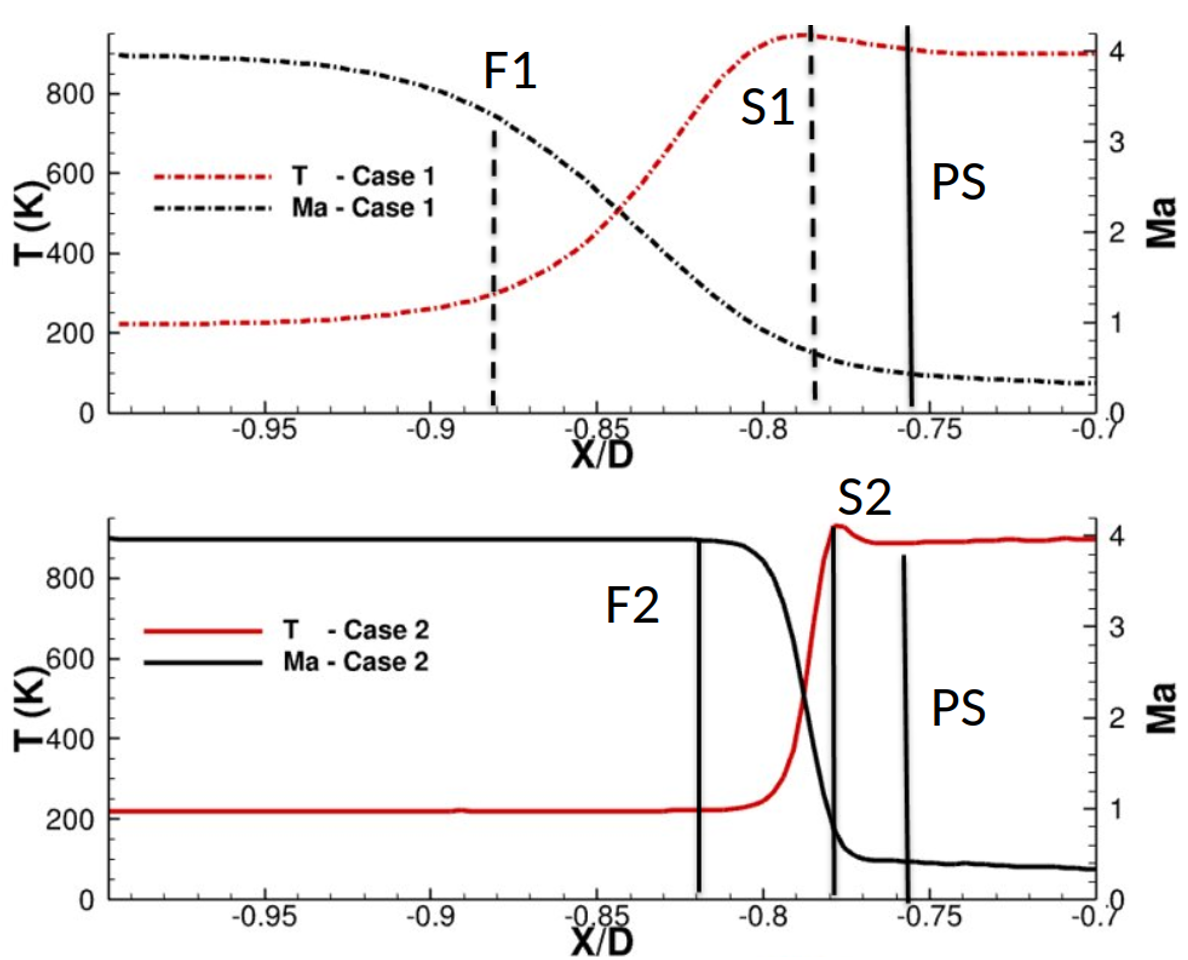}
    \caption{Temperature profiles across the bow shock along the stagnation streamline for Case 1 and Case 2.}
    \label{fig:shock-profiles-temp}
  \end{subfigure}
  \begin{subfigure}[b]{0.46\linewidth}
    \centering
    \includegraphics[width=\linewidth]{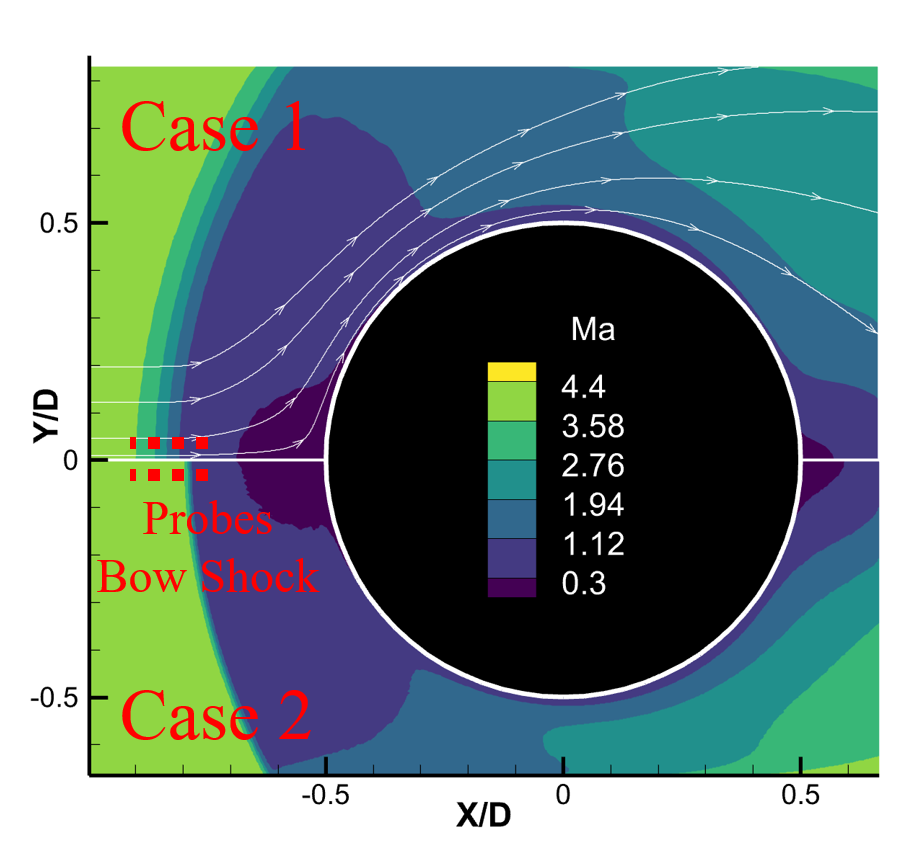}
    \caption{Mach number field with streamlines for Case 1 (top) and Case 2 (bottom), showing probe placement around the shock region.}
    \label{fig:shock-mirror-mach}
  \end{subfigure}
  \caption{Shock structure analysis for Case 1 and Case 2, highlighting temperature profiles and Mach number fields in the stagnation region at \( M = 4 \).}
  \label{fig:shock-structure}
\end{figure}

Figure~\ref{fig:shock-structure} illustrates the shock structure for Cases 1 and 2. Figure~\ref{fig:shock-profiles-temp} shows temperature profiles along the stagnation streamline across the bow shock. In Case 1, the shock exhibits a characteristically diffuse structure due to the elevated Knudsen number (\(Kn_D = 0.0178\)), where the shock thickness becomes comparable to the mean free path. The temperature rise spans a broader region; \( X/D \approx -0.87 \) to \( -0.73 \), approximately \(4 \, \si{\micro\metre}\) or \(7.5\lambda\), reflecting the gradual molecular relaxation processes. In Case 2, the shock is sharper due to the higher number density (lower \(Kn_D = 0.0029\)), with the temperature transition occurring over a narrower region (\( X/D \approx -0.80 \) to \( -0.73 \)). Figure~\ref{fig:shock-mirror-mach} displays the Mach number field with streamlines, highlighting the bow shock's position and structure for both cases.

\begin{table}[H]
  \centering
  \sisetup{
    exponent-product = \times,
    table-number-alignment = center,
    round-mode = figures,
    round-precision = 4
  }
  \caption{Normalized x-locations of probes for two cases}
  \label{tab:probe-locations}
  \begin{minipage}{0.7\textwidth}
    \centering
    \renewcommand{\arraystretch}{1.0}
    \begin{tabular}{
        l
        S[table-format=-1.4]
        S[table-format=-1.4]
      }
      \toprule
       & {Case 1} & {Case 2}                                                                                                                                                                    \\
      \midrule
      Freestream probe (FS) (X/D)
       & -0.8833  & -0.8227                                                                                                                                                                     \\
      Shock probe (S) \ \ \ \ \ \ \ \ (X/D)
       & -0.7858  & -0.7797                                                                                                                                                                     \\
      Post shock probe (PS) (X/D)
       & -0.7718  & -0.7673                                                                                                                                                                     \\
      \midrule
      \multicolumn{3}{l}{\parbox{0.65\textwidth}{\footnotesize Note: Distance is calculated from the cylinder center and normalized by the cylinder diameter ($D = 30 \, \si{\micro\metre}$).}} \\
      \bottomrule
    \end{tabular}
  \end{minipage}
\end{table}

\begin{figure}[H]
  \centering
  \begin{subfigure}[b]{0.48\linewidth}
    \centering
    \includegraphics[width=\linewidth]{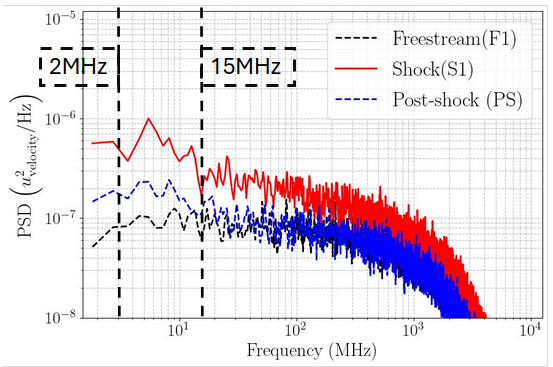}
    \caption{Case 1 (\( N = 2.0 \times 10^{24} \, \mathrm{m}^{-3} \))}
    \label{fig:psd-n2-shock}
  \end{subfigure}\hfill
  \begin{subfigure}[b]{0.48\linewidth}
    \centering
    \includegraphics[width=\linewidth]{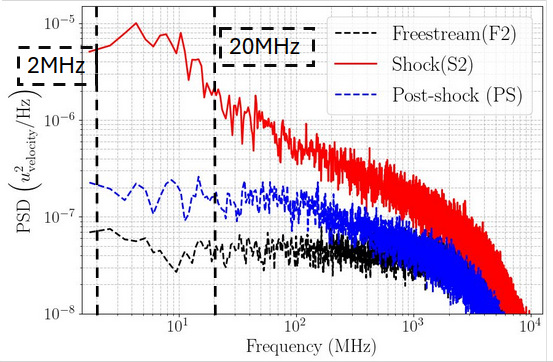}
    \caption{Case 2 (\( N = 1.2 \times 10^{25} \, \mathrm{m}^{-3} \))}
    \label{fig:psd-n12-shock}
  \end{subfigure}
  \caption{PSD analysis of velocity fluctuations in the stagnation region at \( M = 4 \). Three probe locations are compared: Freestream (F1, F2) positioned upstream of the shock in undisturbed supersonic flow; Shock (S1, S2) located within the shock structure at the midpoint of the temperature transition; and Post-Shock (PS) in the subsonic region between the shock and cylinder surface. Distinct spectral peaks between 2--15~MHz (Case 1) and 2--20~MHz (Case 2) appear exclusively at the shock probes, indicating localized oscillatory behavior within the diffuse shock structure.}
  \label{fig:psd-shock}
\end{figure}

The unsteadiness within the bow shock is quantified using Power Spectral Density (PSD) analysis of velocity fluctuations at three probe locations along the stagnation streamline, as shown in Figure~\ref{fig:psd-shock}. The probes are designated as follows: Freestream probes are positioned upstream of the shock in undisturbed flow; Shock probes are located within the shock structure at the midpoint of the temperature rise; and Post-Shock probes are placed in the subsonic region between the shock and the cylinder surface. Subscripts denote the case number (1 or 2). For Case 1 (Figure~\ref{fig:psd-n2-shock}), the PSD at the shock probe (S1) reveals distinct spectral peaks between 2 and 15~MHz, indicating oscillatory behavior absent in both the freestream and post-shock regions. Similarly, for Case 2 (Figure~\ref{fig:psd-n12-shock}), the shock probe (S2) shows peaks between 2 and 20~MHz, again absent in the freestream and post-shock spectra.

These findings differ from prior work by Sawant~\cite{sawantKineticApproachStudying2021a}, who investigated local oscillations in 1D shock structures within a simulated 1-D shock tube flow. In contrast, the present study examines the bow shock ahead of a 30~\si{\micro\metre} diameter cylinder in hypersonic flow (\( M = 4 \)), a geometry introducing curvature and other secondary effects. Sawant’s study suggested that shock oscillation frequencies scale with changes in number density; however, despite a 6-fold increase in number density from Case 1 to Case 2 (\(2.0\times10^{24}\,\mathrm{m}^{-3}\) to \(1.2\times10^{25}\,\mathrm{m}^{-3}\)), the observed frequencies in our bow shock, which are approximately 2–15~MHz in Case 1 and 2–20~MHz in Case 2, do not exhibit a proportional 6-fold increase. This discrepancy may indicate that the unsteadiness in the bow shock is more strongly influenced by its standoff distance, which varies marginally between the cases (as seen in Figure~\ref{fig:shock-structure}), rather than intrinsic molecular interactions within the shock, which would be directly correlated to number density.

It should be noted that probe-based measurement of shock oscillations in DSMC becomes increasingly challenging at higher number densities. As the mean free path decreases, the shock structure sharpens, requiring probe placement with sub-micron precision to capture the oscillatory dynamics within the narrow shock front. The sampling cell must span several mean free paths to ensure statistical convergence, yet remain entirely within the shock structure which is a requirement that becomes mutually exclusive as the shock thickness approaches the sampling resolution. This fundamental limitation motivates the use of spatially-resolved analysis methods, such as Dynamic Mode Decomposition (DMD), which extract coherent oscillatory structures from the entire flowfield without requiring a priori knowledge of their precise spatial location.

\subsection{Wake Instabilities}

\begin{figure}[H]
  \centering
  \begin{subfigure}[b]{0.32\linewidth}
    \centering
    \includegraphics[width=\linewidth]{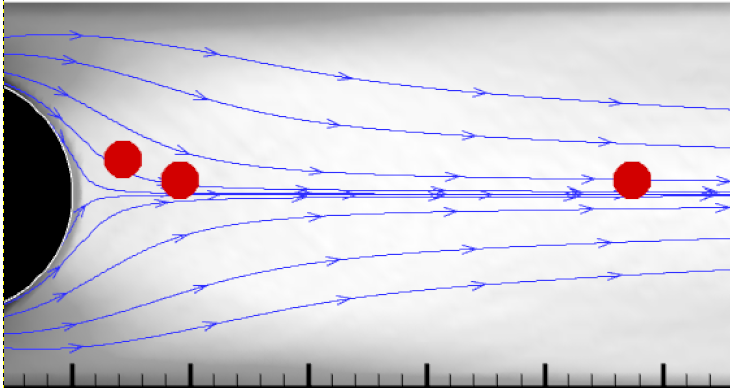}
    \caption{Case 1: \( Re_D = 270 \).}
    \label{fig:N02_layout}
  \end{subfigure}%
  \begin{subfigure}[b]{0.33\linewidth}
    \centering
    \includegraphics[width=\linewidth]{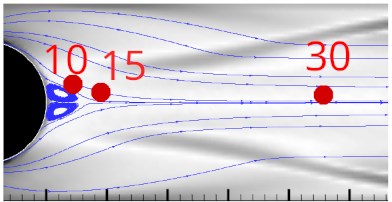}
    \caption{Case 2: \( Re_D = 1620 \).}
    \label{fig:N12_layout}
  \end{subfigure}%
  \begin{subfigure}[b]{0.33\linewidth}
    \centering
    \includegraphics[width=\linewidth]{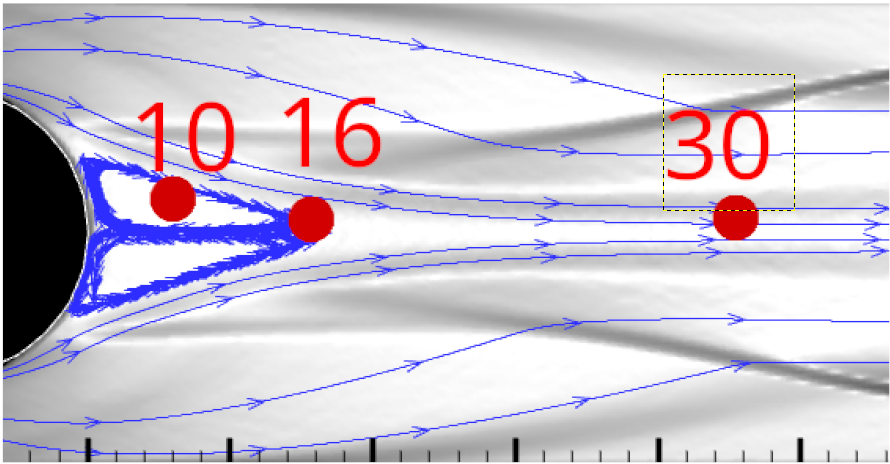}
    \caption{Case 3: \( Re_D = 6750 \).}
    \label{fig:N5_layout}
  \end{subfigure}

  \begin{subfigure}[b]{0.33\linewidth}
    \centering
    \includegraphics[width=\linewidth]{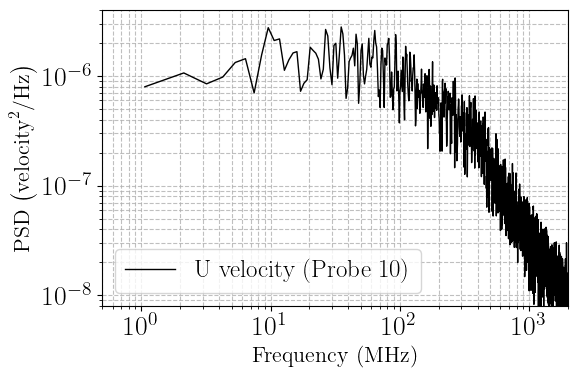}
    \caption{Probe 10, Case 1}
    \label{fig:N02_U10}
  \end{subfigure}%
  \begin{subfigure}[b]{0.33\linewidth}
    \centering
    \includegraphics[width=\linewidth]{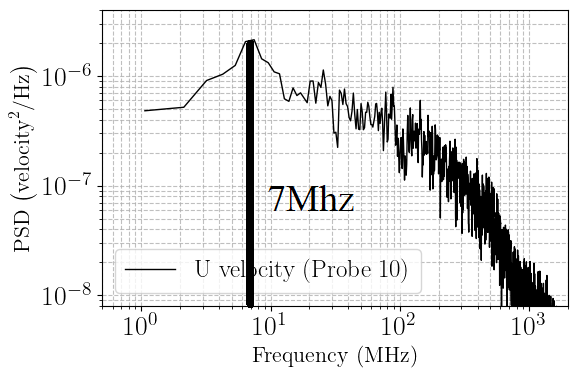}
    \caption{Probe 10, Case 2}
    \label{fig:N12_U10}
  \end{subfigure}%
  \begin{subfigure}[b]{0.33\linewidth}
    \centering
    \includegraphics[width=\linewidth]{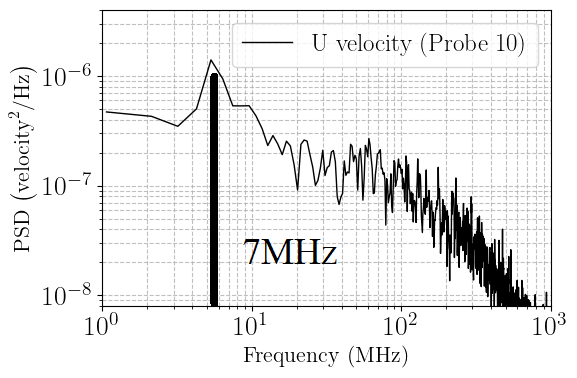}
    \caption{Probe 10, Case 3}
    \label{fig:N5_U10}
  \end{subfigure}

  \begin{subfigure}[b]{0.33\linewidth}
    \centering
    \includegraphics[width=\linewidth]{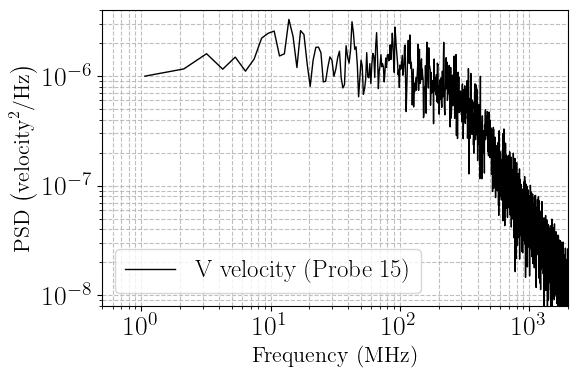}
    \caption{Probe 15, Case 1}
    \label{fig:N02_U15}
  \end{subfigure}%
  \begin{subfigure}[b]{0.33\linewidth}
    \centering
    \includegraphics[width=\linewidth]{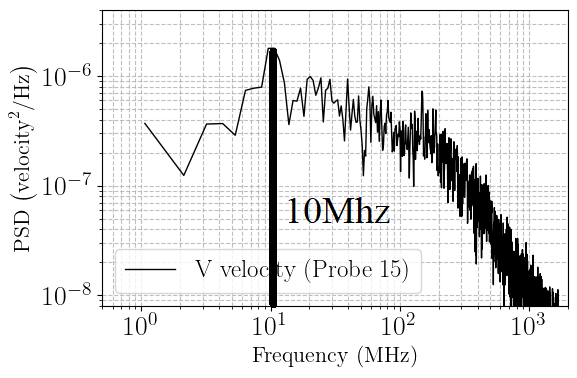}
    \caption{Probe 15, Case 2}
    \label{fig:N12_U15}
  \end{subfigure}%
  \begin{subfigure}[b]{0.33\linewidth}
    \centering
    \includegraphics[width=\linewidth]{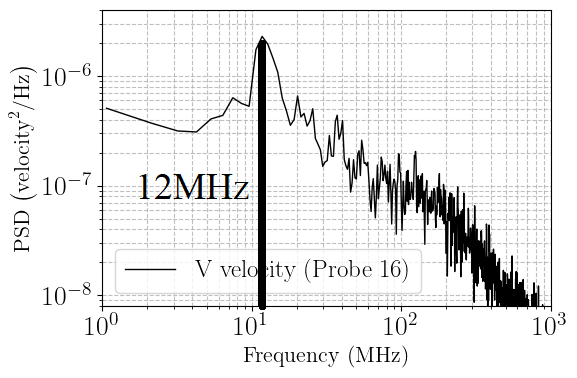}
    \caption{Probe 16, Case 3}
    \label{fig:N5_U16}
  \end{subfigure}

  \begin{subfigure}[b]{0.33\linewidth}
    \centering
    \includegraphics[width=\linewidth]{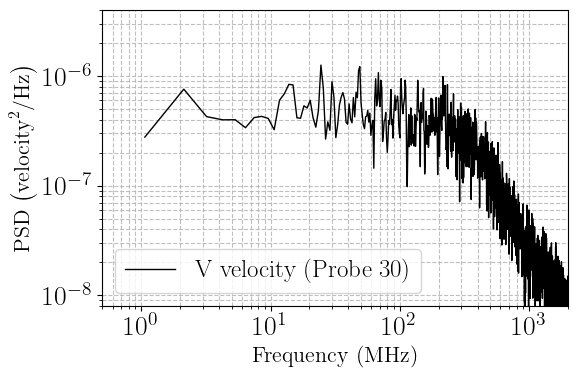}
    \caption{Probe 30, Case 1}
    \label{fig:N02_U30}
  \end{subfigure}%
  \begin{subfigure}[b]{0.33\linewidth}
    \centering
    \includegraphics[width=\linewidth]{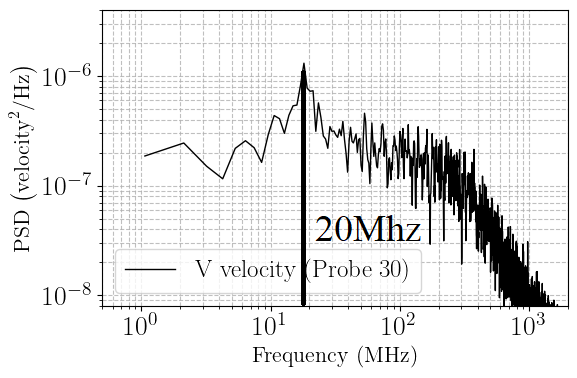}
    \caption{Probe 30, Case 2}
    \label{fig:N12_U30}
  \end{subfigure}%
  \begin{subfigure}[b]{0.33\linewidth}
    \centering
    \includegraphics[width=\linewidth]{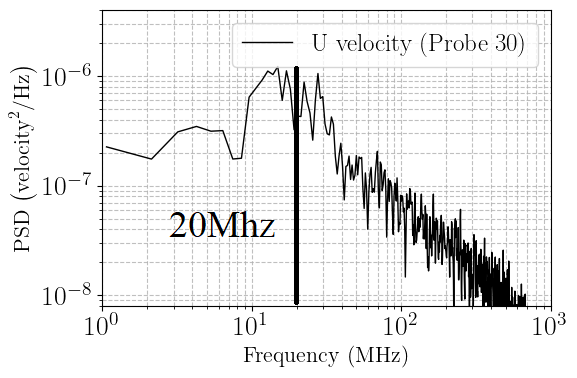}
    \caption{Probe 30,  Case 3.}
    \label{fig:N5_U30}
  \end{subfigure}

  \caption{Probe layouts (top row) and PSD spectra of velocity fluctuations at probes 10, 15 or 16, and 30 for three cases at \( M = 4 \) in the wake region.}
  \label{fig:psd-wake}
\end{figure}

Figure~\ref{fig:psd-wake} shows the probe locations and PSD spectra in the wake region for all three cases. Probes are positioned as follows: probe 10 near the recirculation region and sliplines; probes 15 (Cases 1 and 2) and 16 (Case 3) near the neck; and probe 30 in the far wake.
For Case 1, the PSD spectra (Subfigures~\ref{fig:N02_U10}, \ref{fig:N02_U15}, \ref{fig:N02_U30}) show no distinct frequency peaks at any probe location. This absence of coherent fluctuations is attributed to continuum breakdown in the wake, as discussed above, where rarefaction effects (\( Kn = 0.0178 \)) suppress flow instabilities. Any observed fluctuations are likely due to random kinetic molecular motion rather than organized vortex shedding. In contrast, Cases 2 and 3 exhibit clear frequency peaks, indicating structured unsteadiness. At probe 10 (Subfigures~\ref{fig:N12_U10}, \ref{fig:N5_U10}), both cases show a dominant frequency of 7 MHz, corresponding to a Strouhal number \( St = fD/U_\infty \approx 0.18 \) (with \( D = 30 \, \si{\micro\metre} \), \( U_\infty = 1200 \, \mathrm{m/s} \)), close to the canonical \( St = 0.2 \) for wake oscillations in subsonic flows at similar \( Re_D \).

Further downstream, at probes 15 and 16 (Subfigures~\ref{fig:N12_U15}, \ref{fig:N5_U16}), the dominant frequencies shift to 10 MHz (Case 2) and 12 MHz (Case 3), yielding \( St \approx 0.25 \) and \( St \approx 0.3 \) respectively, aligning with expected vortex shedding frequencies for \( Re_D = 1620 \) and 6750. However, at probe 30 in the far wake (Subfigures~\ref{fig:N12_U30}, \ref{fig:N5_U30}), a higher frequency of 20 MHz emerges in both cases, corresponding to \( St \approx 0.5 \). This elevated Strouhal number suggests a distinct instability mechanism in the far wake of high Mach number flows, differing from subsonic counterparts at similar \( Re_D \). Previous studies \cite{schmidtOscillationsCylinderWakes2015, awasthiSupersonicCylinderWake2022} have reported \( St = 0.4 \) in hypersonic wakes, although in one study \cite{nagataExperimentalInvestigationCompressible2020}, it may have been incorrectly attributed as a harmonic artifact of \( St = 0.2 \). These preliminary results show that the observed perturbation with nearly 20~MHz ($St \approx 0.5$) frequency likely corresponds to a different instability, possibly one driven by shock interactions and wake dynamics around the neck.
Ultimately, this frequency is what will be observed in the far wake region, which matches the results obtained by \cite{schmidtOscillationsCylinderWakes2015}, where they observe a similar frequency and propose an alternate Strouhal definition to describe the phenomenon.
\begin{figure}[H]
  \centering
  \begin{subfigure}[b]{0.70\linewidth}
    \centering
    \includegraphics[width=\linewidth]{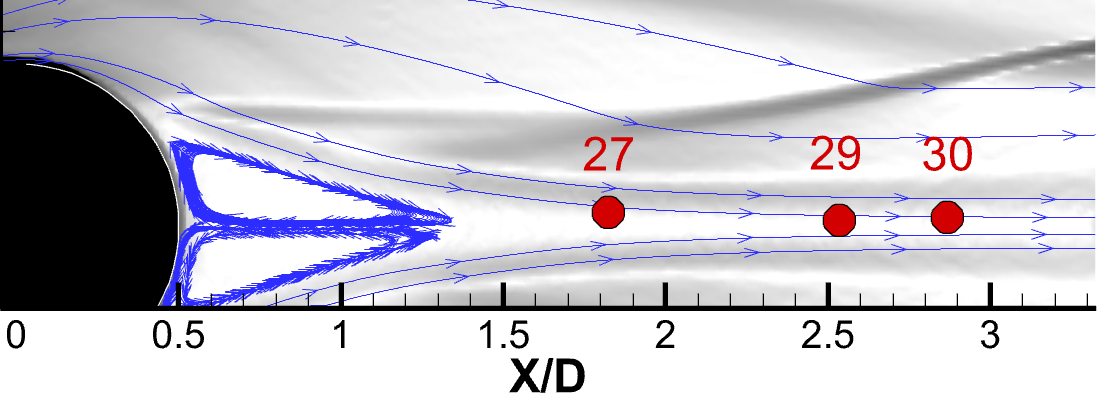}
    \caption{Probes layout, Case-3}
    \label{fig:N02_probes}
  \end{subfigure}%
  \vspace{0.1cm}
  \begin{subfigure}[b]{0.33\linewidth}
    \centering
    \includegraphics[width=\linewidth]{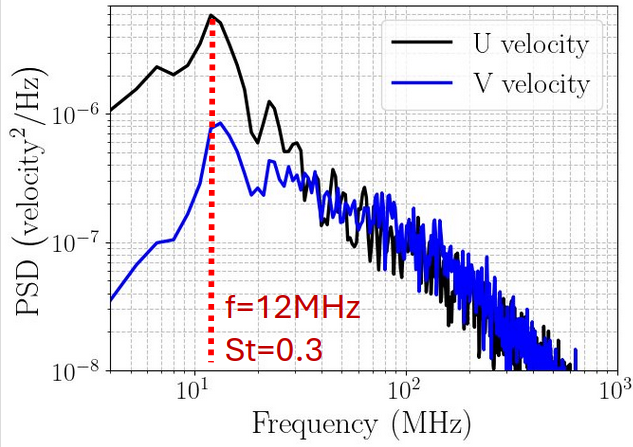}
    \caption{probe 27 Case-3}
    \label{fig:N5_U27}
  \end{subfigure}%
  \begin{subfigure}[b]{0.33\linewidth}
    \centering
    \includegraphics[width=\linewidth]{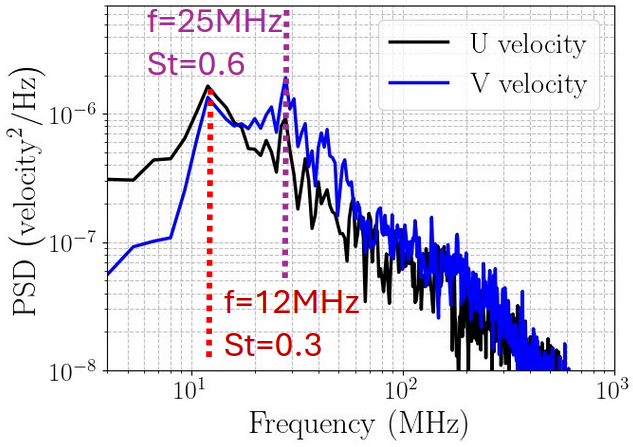}
    \caption{probe 29 Case-3}
    \label{fig:N5_U29}
  \end{subfigure}%
  \begin{subfigure}[b]{0.33\linewidth}
    \centering
    \includegraphics[width=\linewidth]{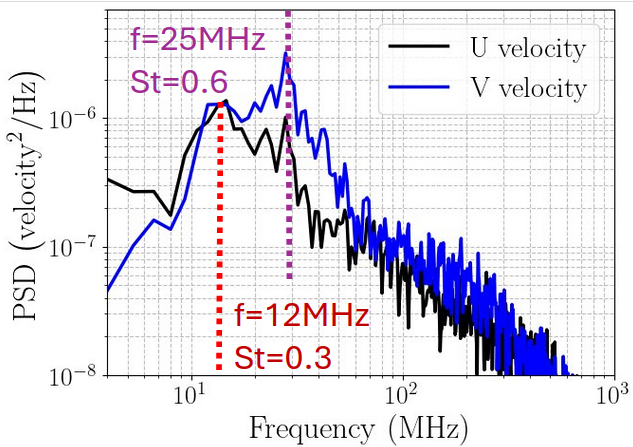}
    \caption{probe 30 Case-3}
    \label{fig:N5_U30_farwake}
  \end{subfigure}
  \caption{Probe layout and PSD of velocity fluctuations in the far wake for Case 3, showing probes 27, 29, and 30 positioned after the neck.}
  \label{fig:FARWAKEmulti}
\end{figure}

To further investigate the far wake unsteadiness, additional probes were placed downstream of the neck for Case 3, as shown in Figure~\ref{fig:FARWAKEmulti}. At probe 27, just downstream of the neck, both the streamwise (U) and transverse (V) velocity components exhibit a peak at approximately 12~MHz (\(St \approx 0.3\)). However, at probe 30 further downstream, a discrepancy emerges: while the 12~MHz peak in the U-velocity spectrum becomes less pronounced, the V-velocity spectrum develops a distinct high-frequency peak at approximately 25~MHz (\(St \approx 0.6\)). This high-frequency mode, appearing exclusively in the transverse velocity component, suggests a new instability mechanism in the far wake that warrants further investigation.

\section{Spatial DMD analysis of the Flowfield}

\subsection{Dynamic Mode Decomposition Framework}

Dynamic Mode Decomposition (DMD) is a data-driven method that decomposes spatio-temporal flowfield data into modes, each associated with a single frequency and growth rate. The method assumes a linear operator \(\mathbf{A}\) advances the system state forward in time: \(\mathbf{q}_{k+1} \approx \mathbf{A}\mathbf{q}_k\), where \(\mathbf{q}_k\) is the flowfield snapshot at timestep \(k\). Through SVD-based rank reduction and eigenvalue decomposition of the reduced operator, DMD extracts spatial modes \(\boldsymbol{\phi}_j\) and corresponding eigenvalues \(\lambda_j\). Each eigenvalue yields a frequency \(\omega_j = \arg(\lambda_j)/\Delta t\) and growth rate \(\sigma_j = \ln|\lambda_j|/\Delta t\), enabling the flowfield reconstruction:
\begin{equation}
  \mathbf{q}(t) \approx \sum_{j=1}^{r} b_j \boldsymbol{\phi}_j e^{(\sigma_j + i\omega_j) t}
  \label{eq:dmd-reconstruction}
\end{equation}
where \(b_j\) are mode amplitudes and \(r\) is the truncation rank. Each term represents a spatially coherent structure oscillating at a single frequency.

Unlike Proper Orthogonal Decomposition (POD), which enforces spatial orthogonality between modes, DMD separates structures by frequency. POD requires \(\langle \boldsymbol{\psi}_i, \boldsymbol{\psi}_j \rangle = 0\) for \(i \neq j\), meaning the spatial inner product of any two distinct modes must vanish. This is a mathematical constraint imposed by the algorithm, not a physical property of the flow. There is no reason why coherent structures oscillating at different frequencies should have zero spatial correlation. Consequently, if two instabilities at different frequencies have overlapping spatial support, POD modes will each contain contributions from both, mixing distinct phenomena. DMD assigns each frequency to a separate mode regardless of spatial overlap. The truncation rank \(r\) is selected based on the singular value decomposition (SVD), where the number of DMD modes is chosen such that modes above the noise floor are retained to capture physical dynamics without overfitting.
When multiple frequencies coexist within an analysis domain, the spatial extent of each oscillatory structure affects its representation in the DMD spectrum. A signal occupying a larger spatial region will dominate the mode amplitudes even if its local amplitude is comparable to other signals. The analysis domain must therefore be chosen to isolate the dynamics of interest.

\begin{figure}[H]
  \centering
  \includegraphics[width=0.8\linewidth]{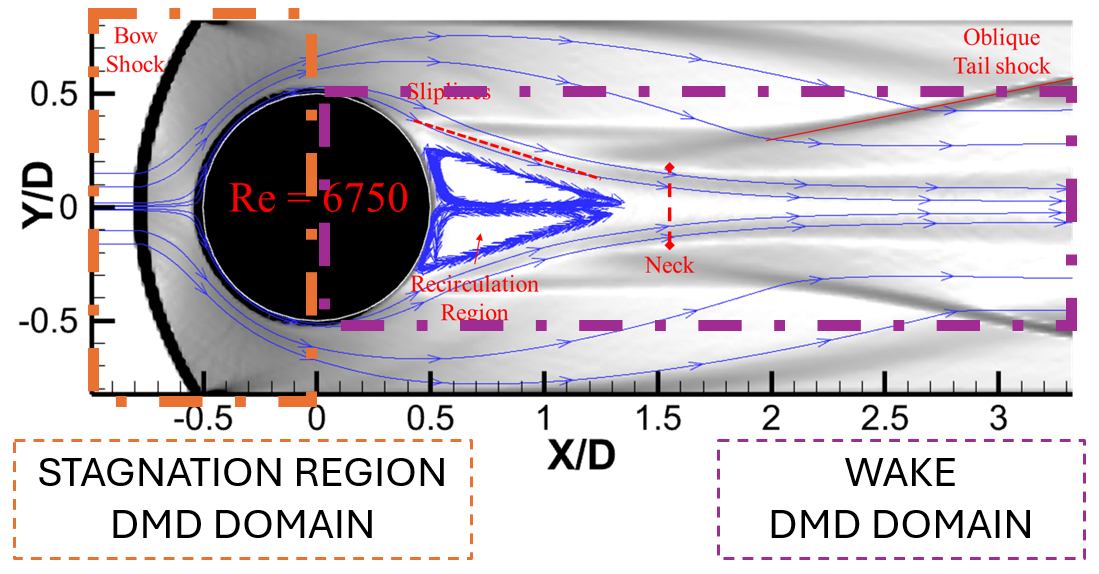}
  \caption{Analysis domains for DMD: stagnation region (left) and wake region (right).}
  \label{fig:dmd-domains}
\end{figure}

For this study, the stagnation region and wake are analyzed separately, as shown in Figure~\ref{fig:dmd-domains}. The stagnation domain extends from \(X/D = -1\) to \(X/D = 0\) which is the cylinder center, and spans \(Y/D = \pm 0.8\) in the transverse direction. The wake analysis domain spans from the cylinder center to \(X/D = 3.3\) downstream with \(Y/D = \pm 0.5\) transversely. This separation prevents the distinct frequency content of each region from interfering in the decomposition. Based on the singular value distributions, 5 DMD modes are retained for the stagnation region and 8 for the wake. All other parameters are kept identical between the DMD stagnation and wake region analyses.

\begin{figure}[H]
  \centering
  \begin{subfigure}[b]{0.45\linewidth}
    \centering
    \includegraphics[width=\linewidth]{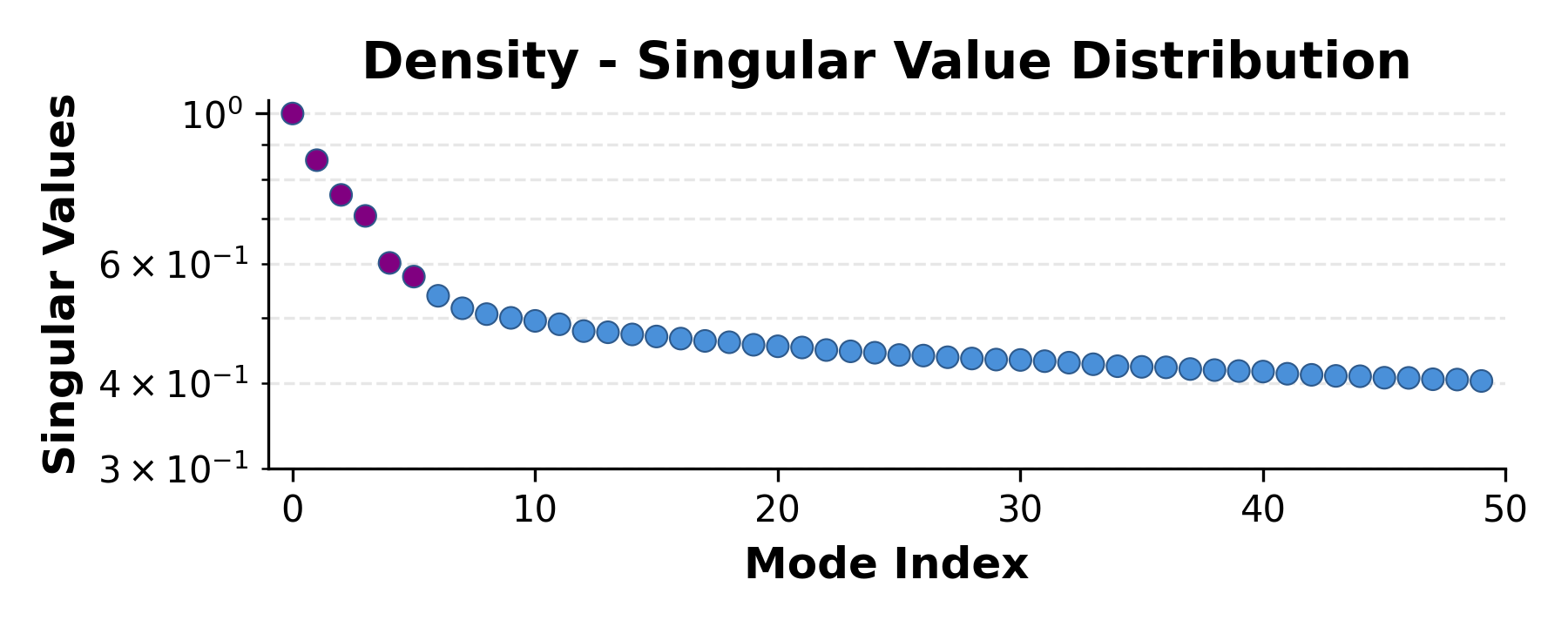}
    \caption{Singular value distribution (number density)}
    \label{fig:svd-ndensity}
  \end{subfigure}%
  \hfill
  \begin{subfigure}[b]{0.45\linewidth}
    \centering
    \includegraphics[width=\linewidth]{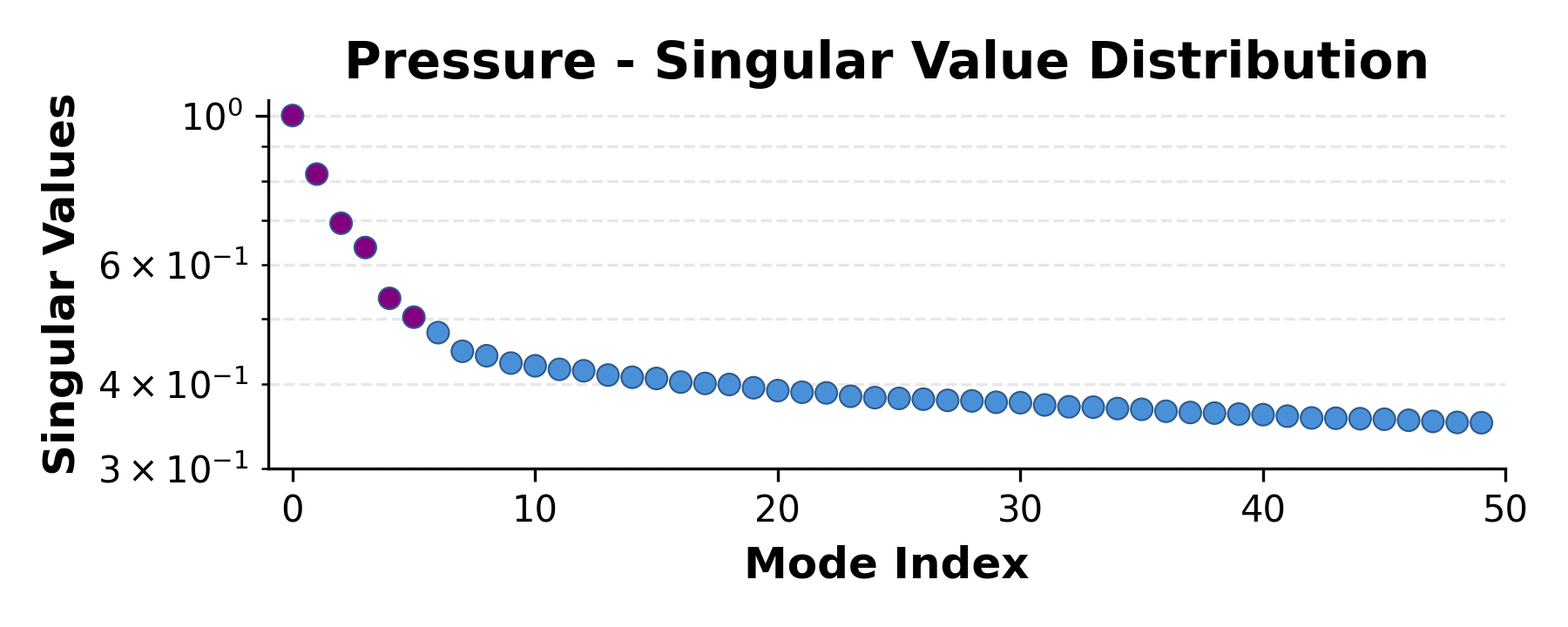}
    \caption{Singular value distribution (pressure)}
    \label{fig:svd-pressure}
  \end{subfigure}

  \vspace{0.5em}

  \begin{subfigure}[b]{0.45\linewidth}
    \centering
    \includegraphics[width=\linewidth]{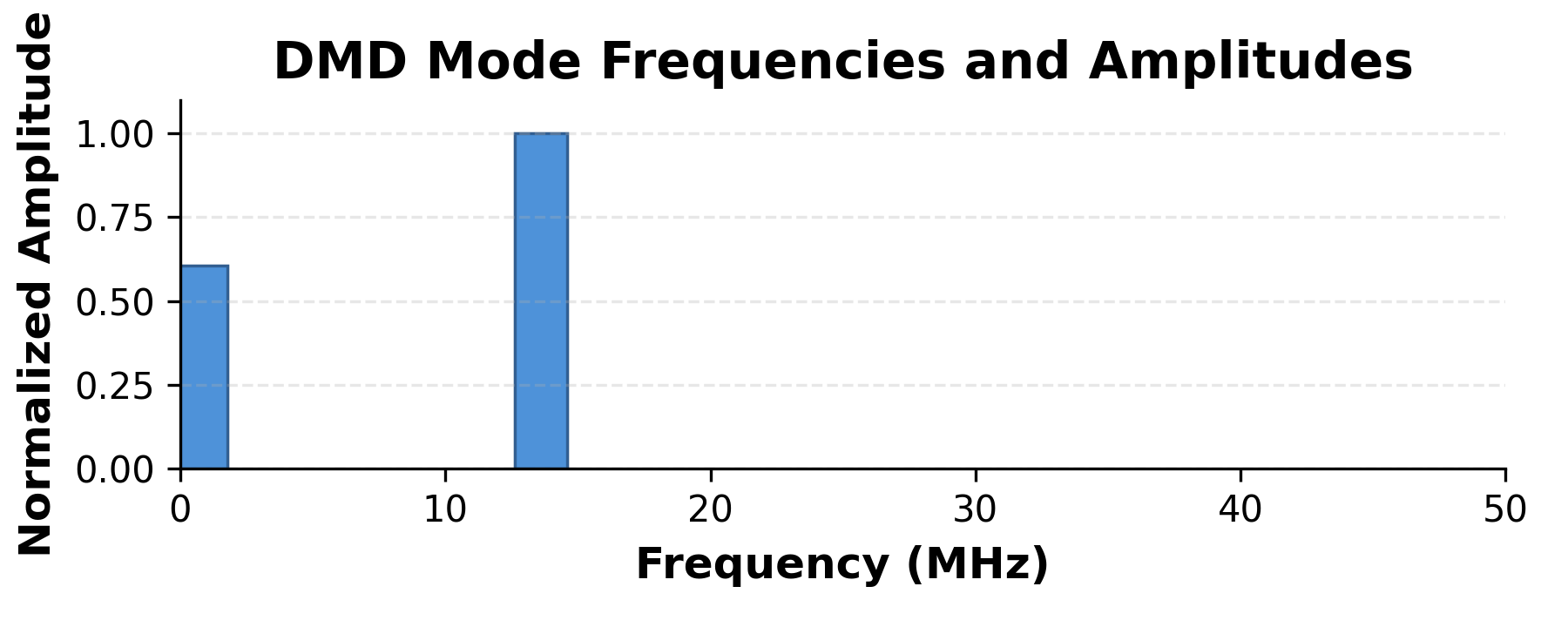}
    \caption{DMD mode frequencies and amplitudes (density)}
    \label{fig:amp-ndensity}
  \end{subfigure}%
  \hfill
  \begin{subfigure}[b]{0.45\linewidth}
    \centering
    \includegraphics[width=\linewidth]{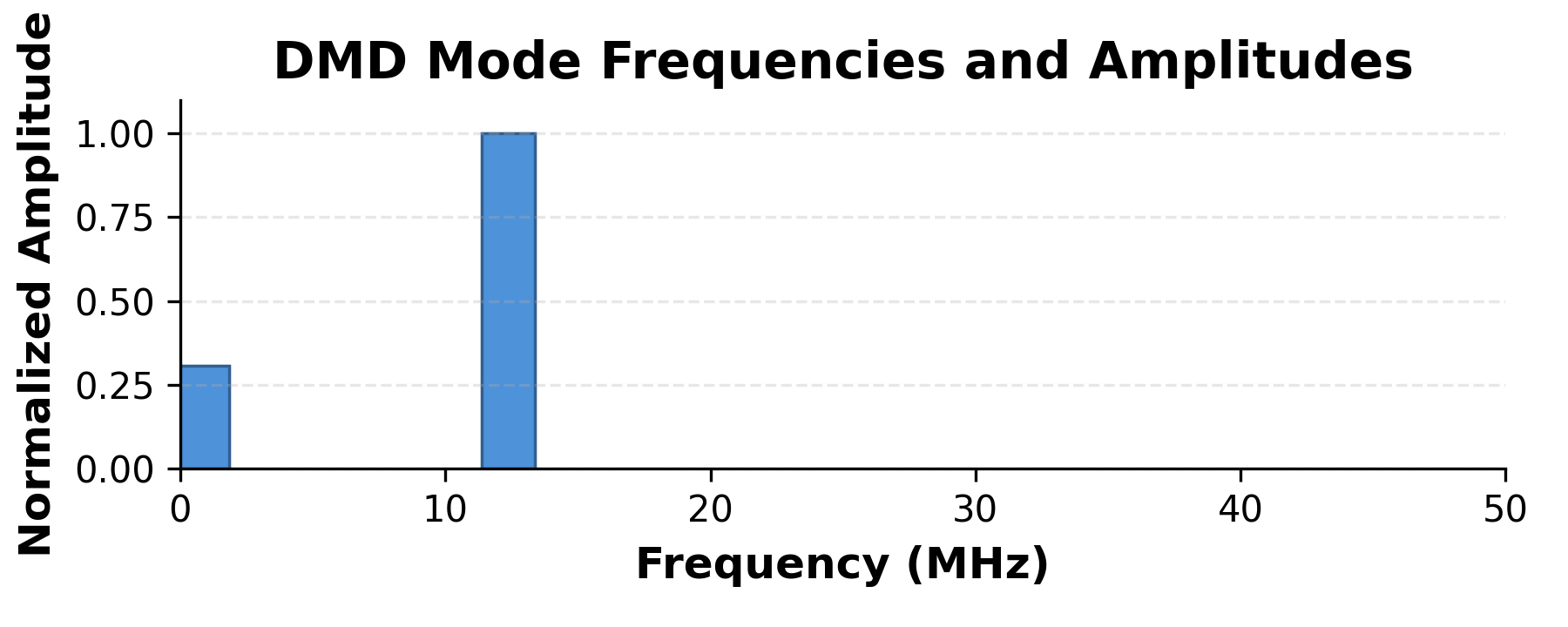}
    \caption{DMD mode frequencies and amplitudes (pressure)}
    \label{fig:amp-pressure}
  \end{subfigure}

  \vspace{0.5em}

  \begin{subfigure}[b]{0.49\linewidth}
    \centering
    \includegraphics[width=\linewidth]{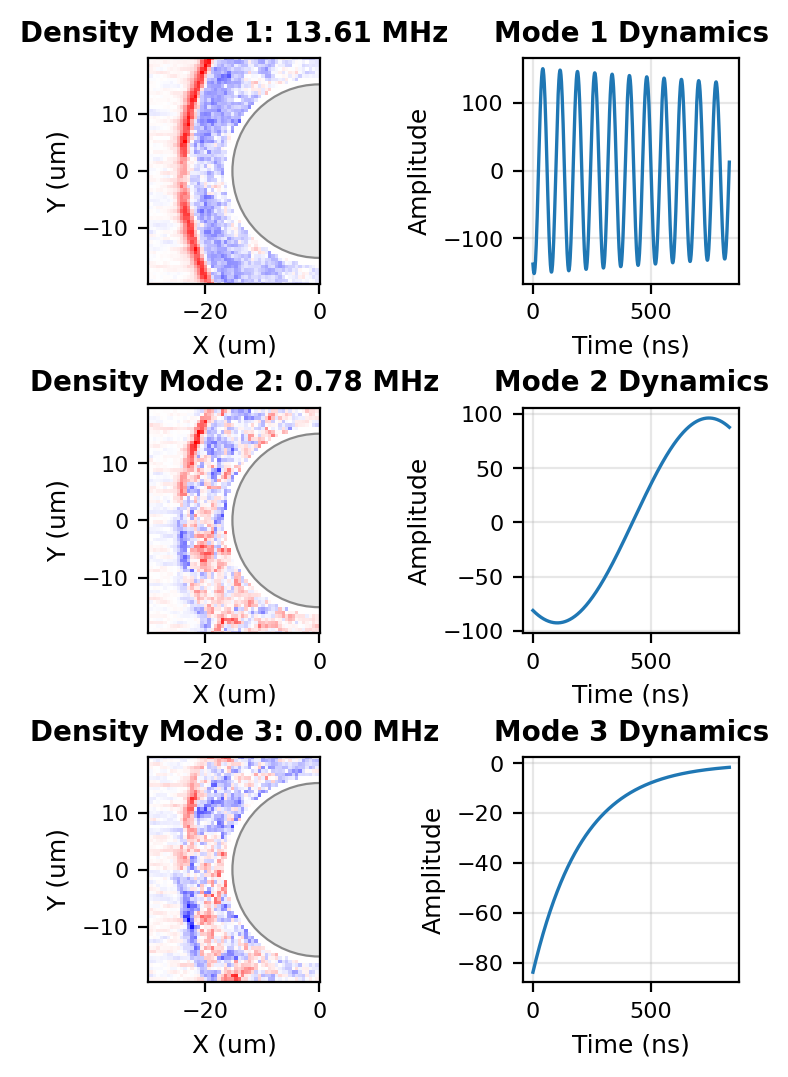}
    \caption{Spatial modes 1--3 with temporal dynamics (density)}
    \label{fig:modes-ndensity}
  \end{subfigure}%
  \hfill
  \begin{subfigure}[b]{0.49\linewidth}
    \centering
    \includegraphics[width=\linewidth]{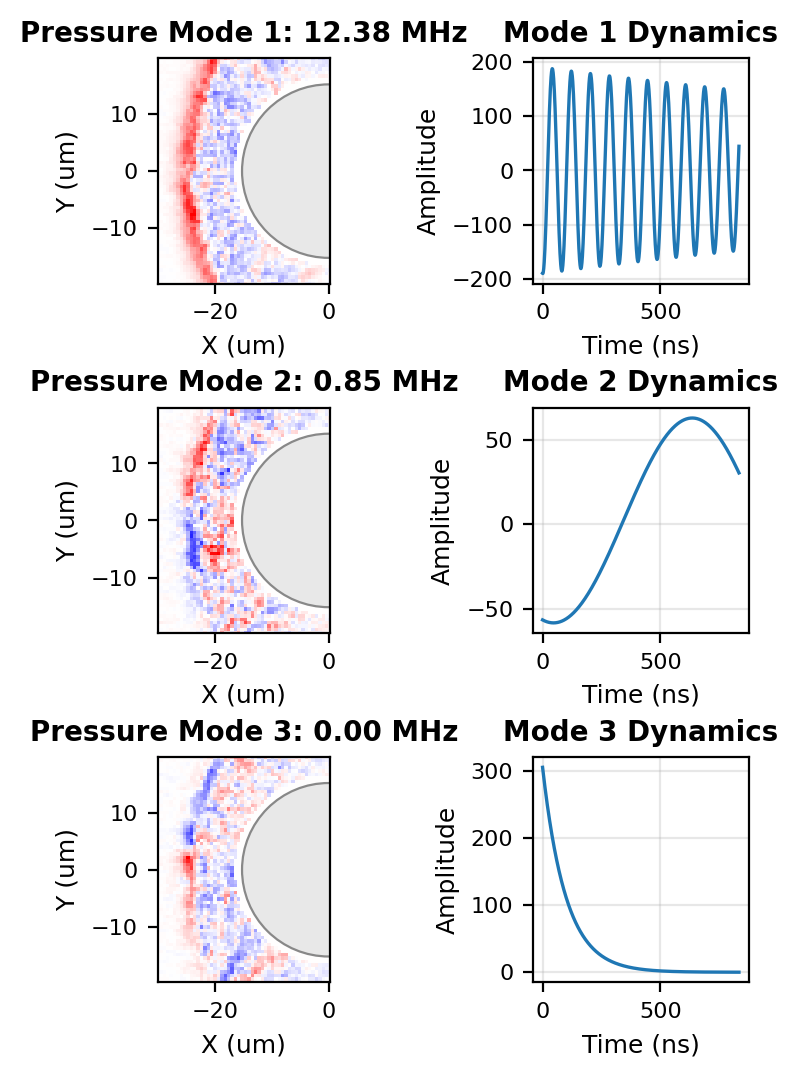}
    \caption{Spatial modes 1--3 with temporal dynamics (pressure)}
    \label{fig:modes-pressure}
  \end{subfigure}

  \caption{DMD analysis of the stagnation region for Case 1 (\(Re_D = 270\), \(Kn_D = 0.0178\)). Left column: number density field; right column: pressure field. (a,b) Singular value distributions confirming low-rank structure. (c,d) Normalized mode amplitudes versus frequency; the dominant mode at $\sim$13~MHz corresponds to bow shock pulsation. (e,f) Spatial structure and temporal evolution of modes 1--3; Mode 1 captures the oscillatory shock dynamics, Mode 2 contains residual low-frequency content, and Mode 3 (zero frequency) represents transient decay.}
  \label{fig:dmd-case1-summary}
\end{figure}

Figure~\ref{fig:dmd-case1-summary} presents the DMD analysis for the stagnation region of Case 1, with number density and pressure fields analyzed independently. The singular value distributions (Figures~\ref{fig:svd-ndensity} and~\ref{fig:svd-pressure}) show that the data is low-rank, with a small number of modes capturing the dominant dynamics. Since DMD modes appear as complex conjugate pairs, 6 retained modes correspond to at most 3 distinct frequencies.

The mode amplitude spectra (Figures~\ref{fig:amp-ndensity} and~\ref{fig:amp-pressure}) show the dominant mode at 13.6~MHz for number density and 12.4~MHz for pressure. These frequencies are consistent with each other and correspond to pulsation of the bow shock and stagnation region. They also fall within the 2--15~MHz band observed in the probe-based PSD analysis (Figure~\ref{fig:psd-n2-shock}), confirming agreement between the two methods.

The spatial structure of each mode is shown in Figures~\ref{fig:modes-ndensity} and~\ref{fig:modes-pressure}. Mode 1, the dominant oscillatory mode at approximately 13~MHz, shows fluctuations localized at the bow shock front with alternating positive and negative values, consistent with shock position oscillations. Mode 2 captures lower-frequency content near 0.8~MHz. Mode 3 appears at zero frequency, corresponding to residual transient drift that has not fully decayed. The temporal dynamics confirm these interpretations: Mode 1 oscillates with constant amplitude, while Mode 3 shows monotonic decay. The 13~MHz pulsation is the dominant unsteady signal in the stagnation region.

\subsection{Spatial DMD Analysis of the Stagnation Region}
\begin{figure}[H]
  \centering
  \begin{subfigure}[b]{0.33\linewidth}
    \centering
    \includegraphics[width=\linewidth,trim=230 0 145 0,clip]{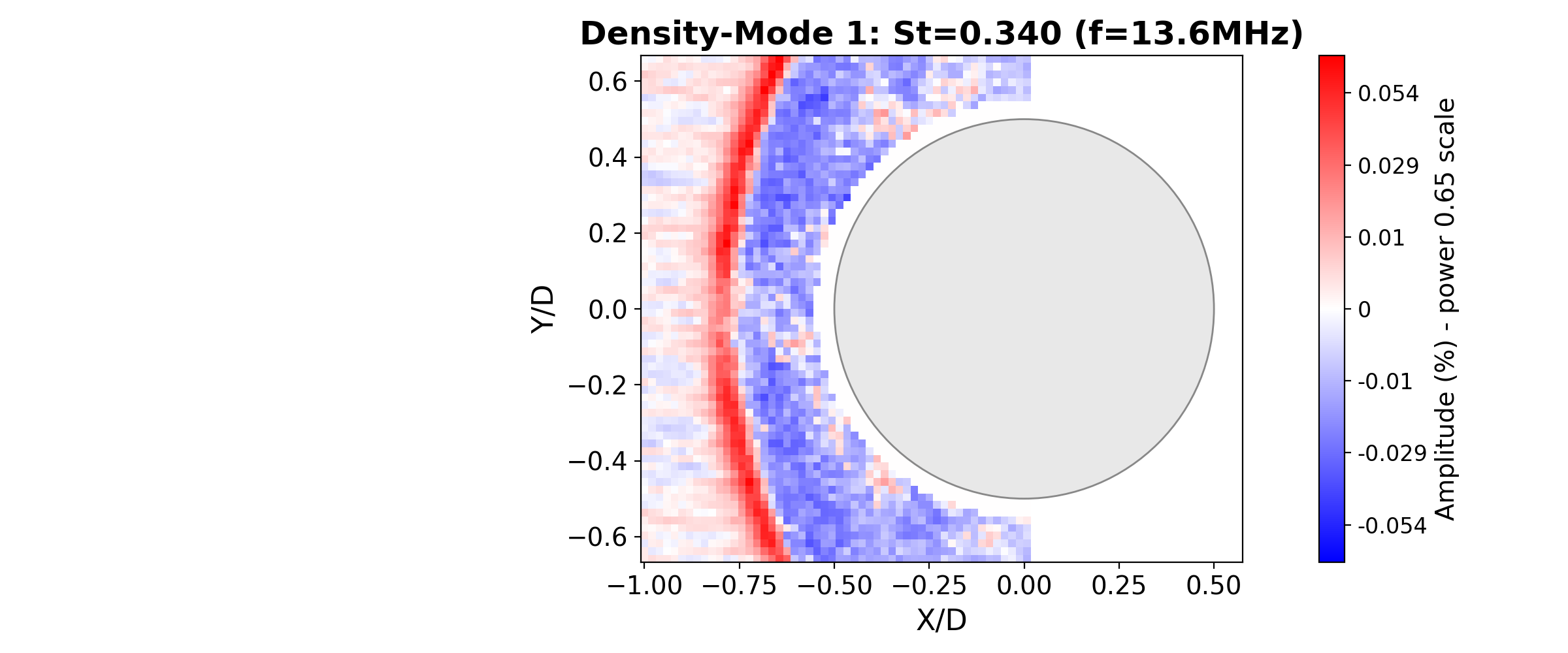}
    \caption{Case 1: $St = 0.340$ (\(Re_D = 270\))}
    \label{fig:dmd-case1-stag}
  \end{subfigure}%
  \begin{subfigure}[b]{0.33\linewidth}
    \centering
    \includegraphics[width=\linewidth,trim=230 0 145 0,clip]{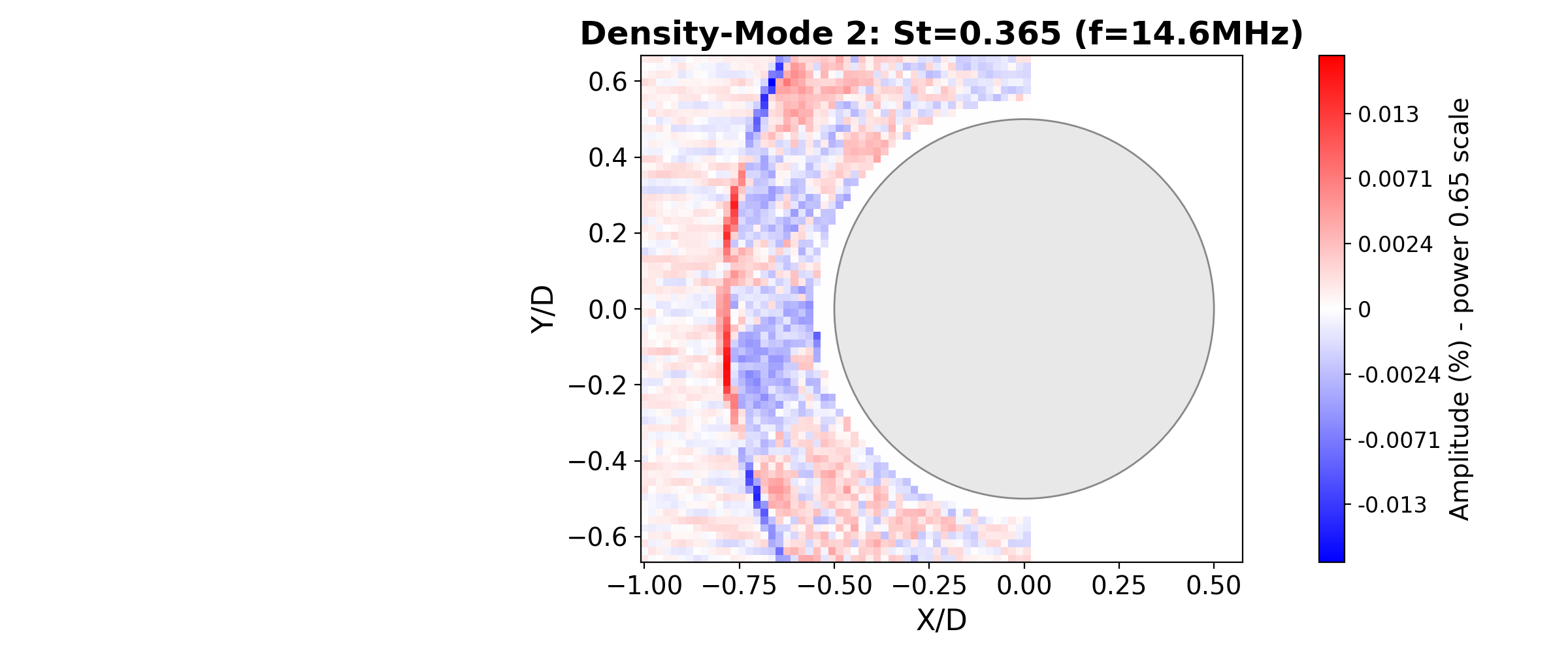}
    \caption{Case 2: $St = 0.365$ (\(Re_D = 1620\))}
    \label{fig:dmd-case2-stag}
  \end{subfigure}%
  \begin{subfigure}[b]{0.33\linewidth}
    \centering
    \includegraphics[width=\linewidth,trim=230 0 145 0,clip]{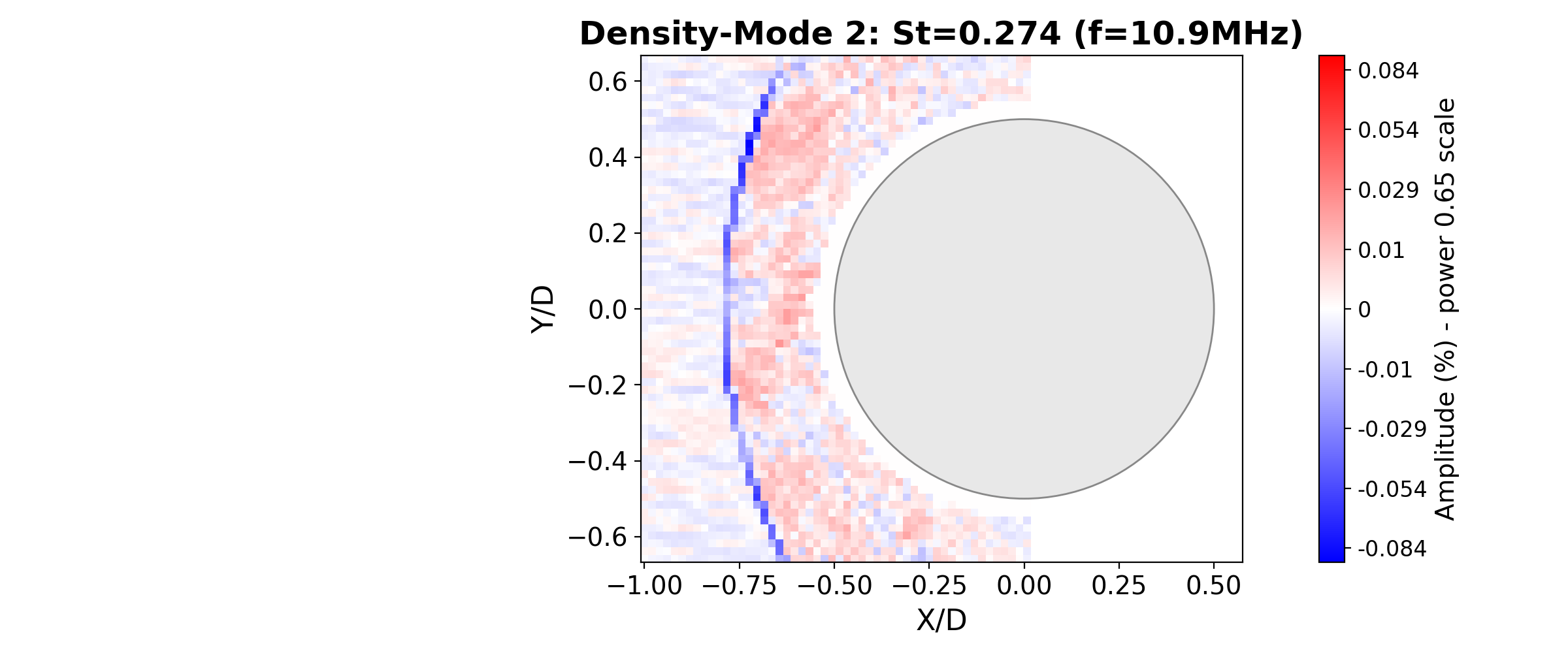}
    \caption{Case 3: $St = 0.274$ (\(Re_D = 6750\))}
    \label{fig:dmd-case3-stag}
  \end{subfigure}
  \caption{Dominant oscillatory DMD modes in the stagnation region for all three cases. Despite a 24-fold increase in number density from Case 1 to Case 3, the spatial structure of the pulsating shock mode remains qualitatively similar, and the dominant frequencies (10--15~MHz) vary only modestly and cluster around $St \approx 0.3$. This insensitivity suggests that the oscillation timescale is likely governed by the shock standoff distance.}
  \label{fig:dmd-bowshock}
\end{figure}

Figure~\ref{fig:dmd-bowshock} compares the dominant oscillatory DMD mode across all three cases, spanning a 24-fold range in number density. Results are presented in non-dimensional notation as \( St = fD/U_\infty \), where \( D = \SI{30}{\micro\metre} \) is the cylinder diameter and \( U_\infty = \SI{1200}{\metre\per\second} \) is the freestream velocity.

The spatial structure of the pulsating mode is similar in each case: alternating positive and negative fluctuations localized at the bow shock front, indicative of shock position oscillations. The extracted frequencies are $St = 0.340$ (Case 1), $St = 0.365$ (Case 2), and $St = 0.274$ (Case 3), all within the same order of magnitude. The weak dependence on number density is notable. If the oscillation frequency were set by molecular collision timescales, one would expect it to scale proportionally with density. The observed near-constant frequency instead suggests that the shock standoff distance is the controlling length scale. The captured frequencies consistently cluster around $St \approx 0.3$ (10--15~MHz).
These findings suggest that bow shock pulsation may occur at any scale where a detached shock forms ahead of a blunt body. The absolute frequency would scale inversely with standoff distance, ranging from kilohertz for macroscale geometries to megahertz for the microscale cylinder studied here.

At higher Reynolds numbers, the stagnation region appears to exhibit coupling with the wake. In Cases 2 and 3, secondary DMD modes at frequencies coinciding with wake instabilities are observed in the stagnation domain. These modes can be distinguished from the symmetric pulsation mode by their asymmetric spatial structure, with opposite-sign fluctuations above and below the centerline. This asymmetry is suggestive of vertical oscillation of the bow shock position, potentially driven by the alternating pressure field associated with vortex shedding in the wake. The presence of wake-associated frequencies in the upstream stagnation region suggests that the flowfield may be globally coupled at higher Reynolds numbers.

\subsection{DMD analysis of the wake}

The wake region is analyzed separately from the stagnation domain to isolate wake-specific instabilities. Results are non-dimensionalized with canonical notation, where time is normalized as \( \tau = t / (D/U_\infty) \), representing convective time units. DMD is applied to both streamwise (U) and transverse (V) velocity fields, with 8 modes retained based on the singular value spectrum. Case 1 is excluded from this analysis due to the absence of a recirculation region and coherent wake instabilities.
\begin{figure}[H]
  \centering
  \begin{subfigure}[b]{0.48\linewidth}
    \centering
    \includegraphics[width=\linewidth]{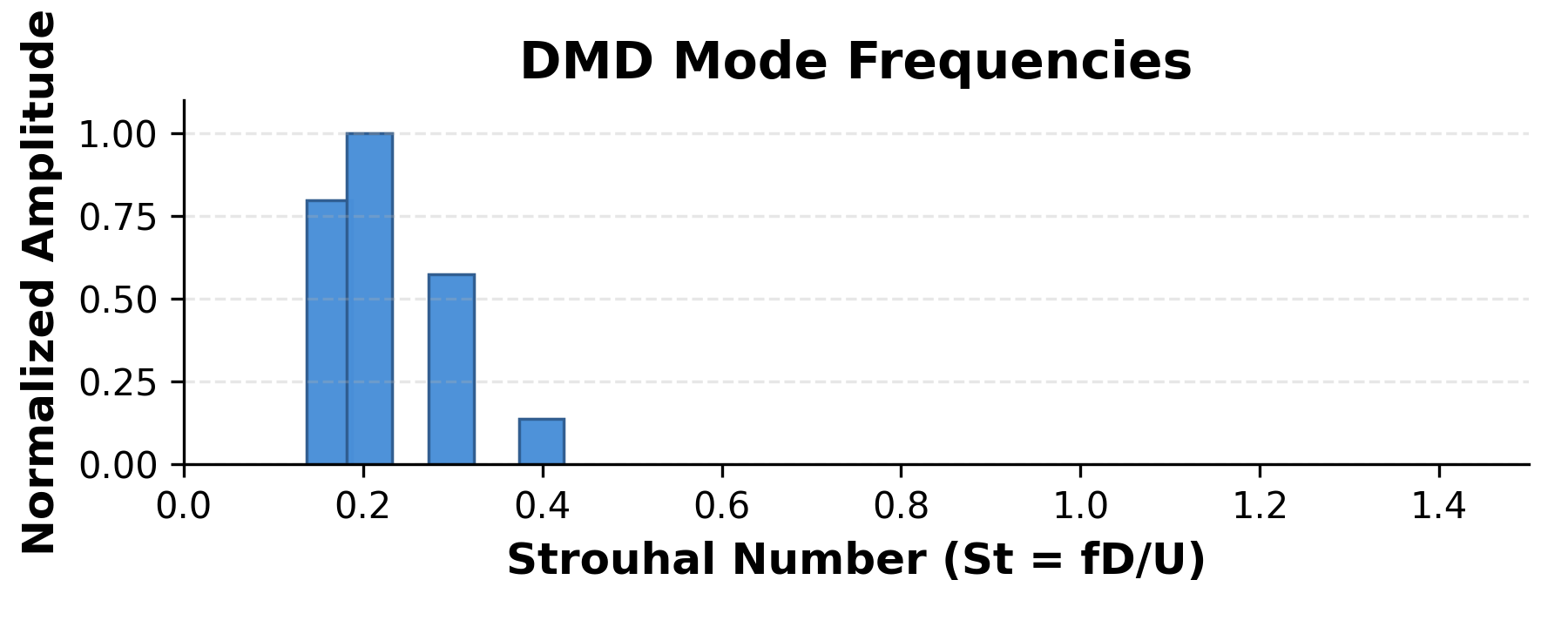}
    \caption{U-velocity: mode amplitudes}
    \label{fig:dmd-case2-u-freq}
  \end{subfigure}%
  \hfill
  \begin{subfigure}[b]{0.48\linewidth}
    \centering
    \includegraphics[width=\linewidth]{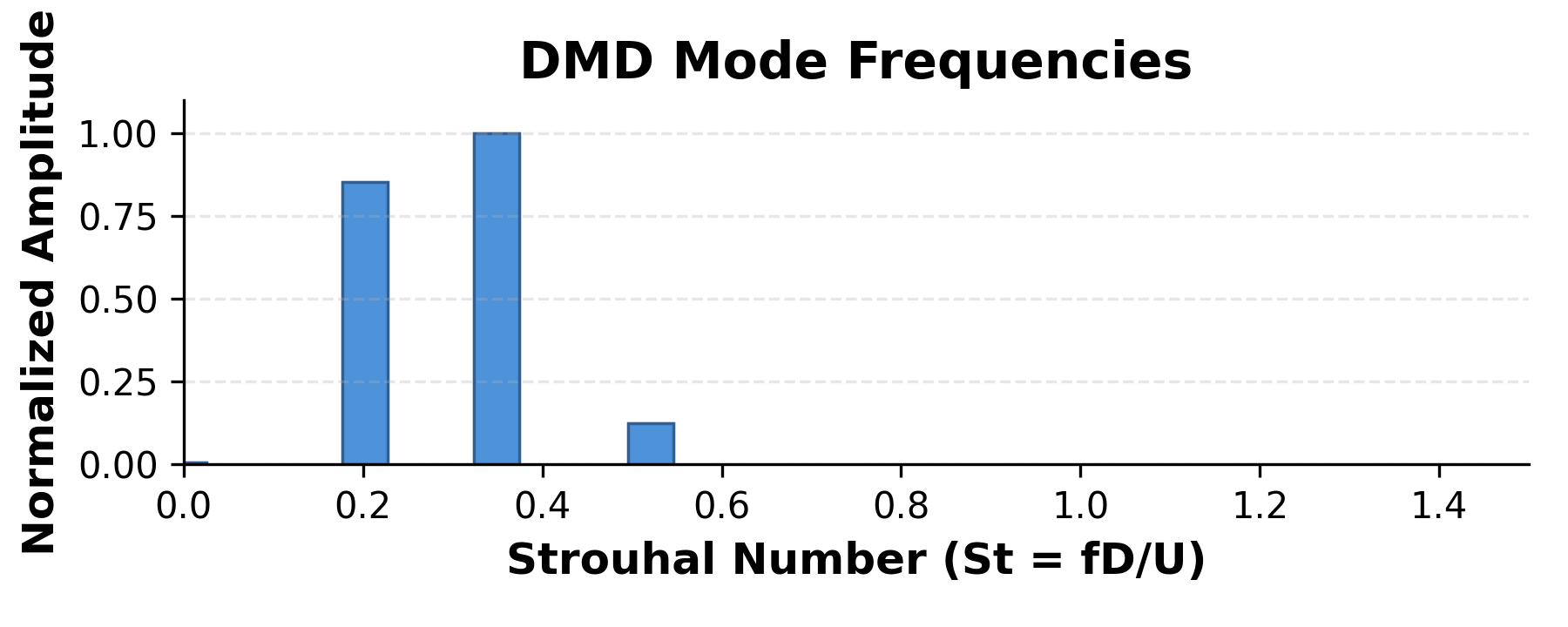}
    \caption{V-velocity: mode amplitudes}
    \label{fig:dmd-case2-v-freq}
  \end{subfigure}
  \vspace{0.5em}
  \begin{subfigure}[b]{0.48\linewidth}
    \centering
    \includegraphics[width=\linewidth]{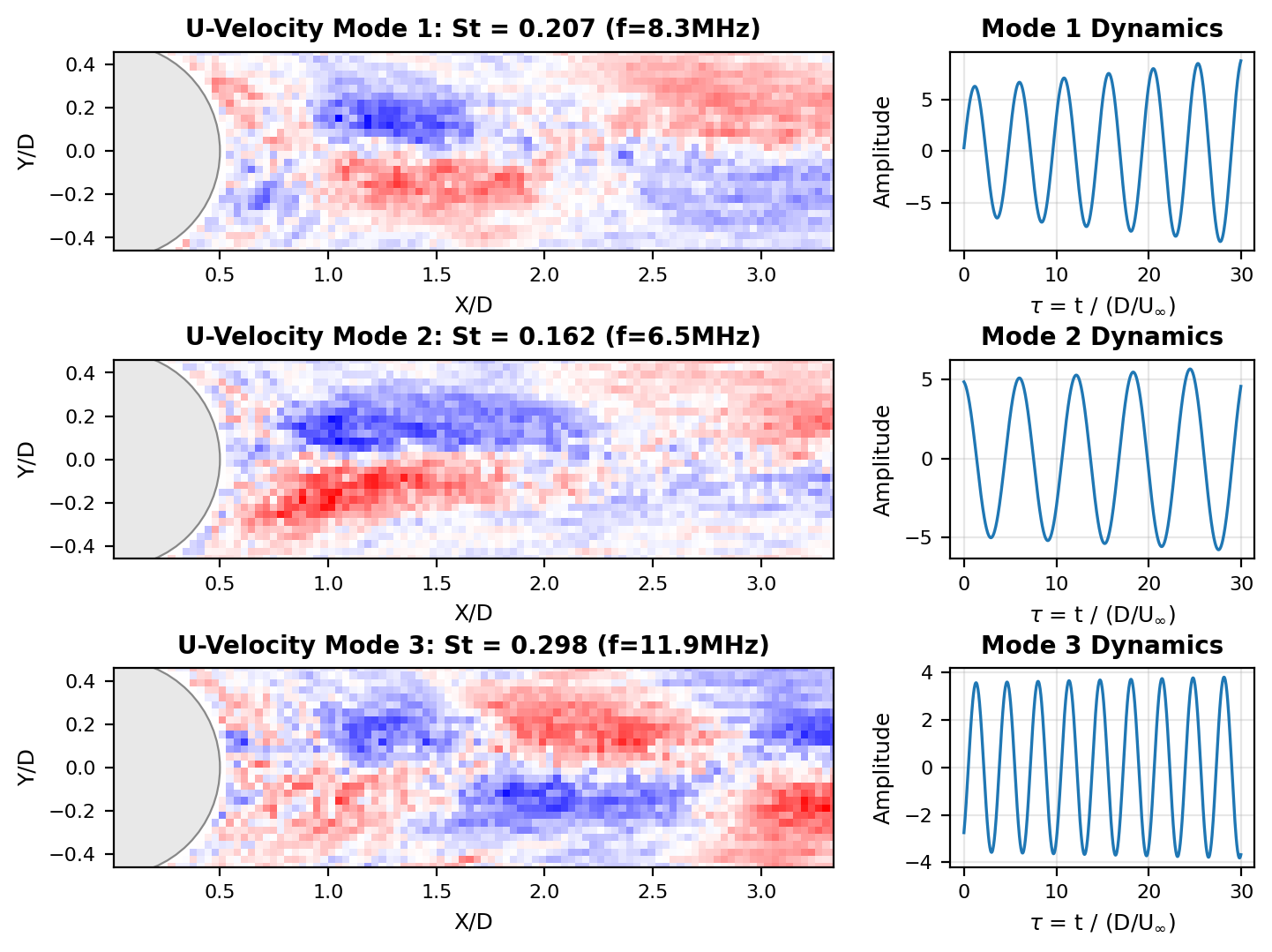}
    \caption{U-velocity modes: mainly the vortex shedding modes around \(St \approx 0.2\)}
    \label{fig:dmd-case2-u-modes}
  \end{subfigure}%
  \hfill
  \begin{subfigure}[b]{0.48\linewidth}
    \centering
    \includegraphics[width=\linewidth]{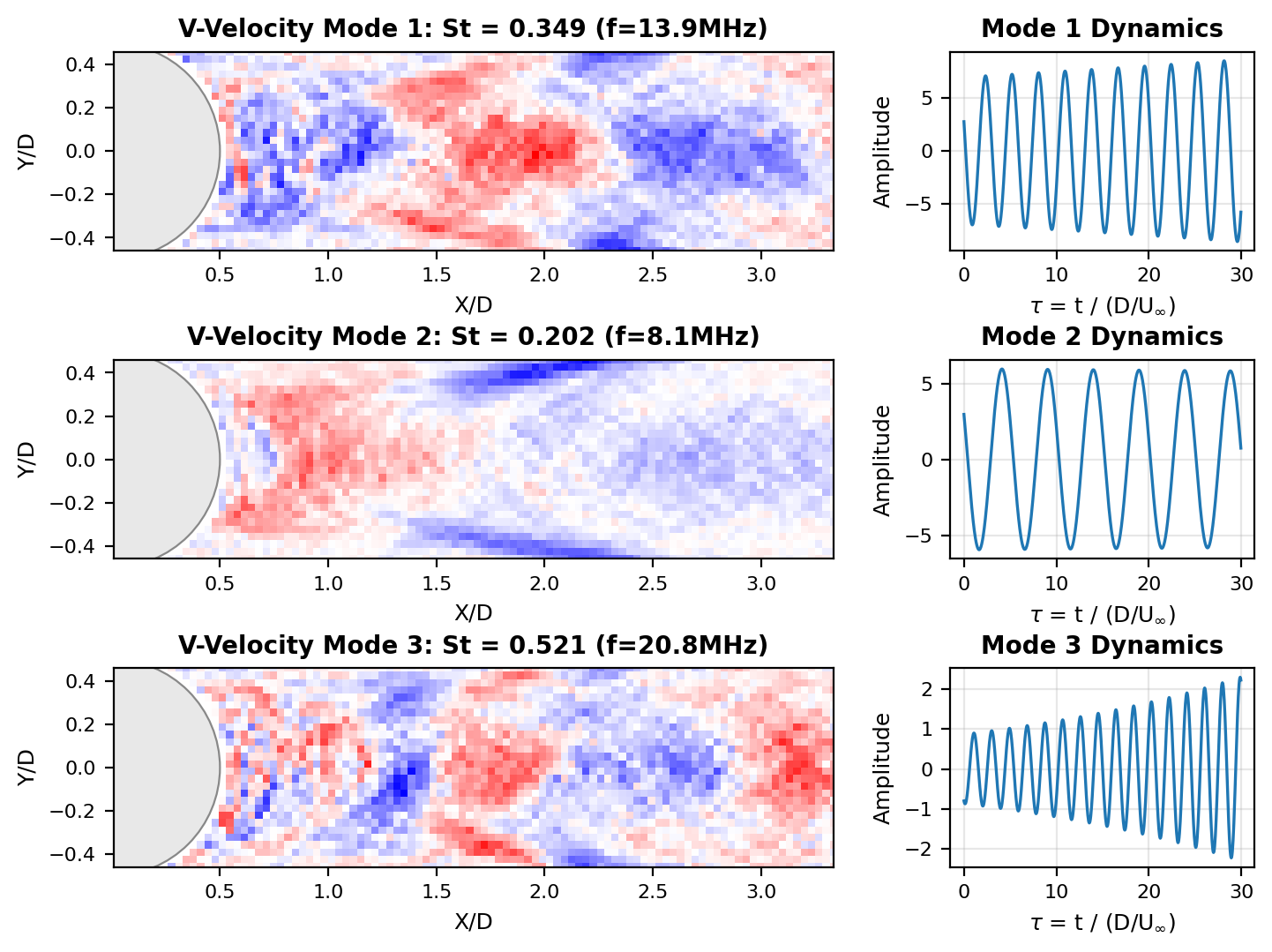}
    \caption{V-velocity modes: shedding (Mode 2) and emerging \(St \approx 0.5\) mode (Mode 3)}
    \label{fig:dmd-case2-v-modes}
  \end{subfigure}
  \caption{Wake DMD analysis for Case 2 (\(Re_D = 1620\)). Left column: streamwise (U) velocity. Right column: transverse (V) velocity. Top row: normalized mode amplitudes versus frequency. Bottom row: spatial structure and temporal dynamics of modes 1--3. The U-velocity field is dominated by the canonical shedding mode at \(St \approx 0.2\), while the V-velocity field shows an emerging secondary mode at \(St \approx 0.5\).}
  \label{fig:dmd-case2-wakeNondim}
\end{figure}

Figure~\ref{fig:dmd-case2-wakeNondim} presents the wake DMD results for Case 2 (\(Re_D = 1620\)). The U-velocity amplitude spectrum (Figure~\ref{fig:dmd-case2-u-freq}) shows multiple modes clustering around \(St \approx 0.2\) (7--12~MHz). This frequency band captures the canonical vortex shedding instability, consistent with the earlier PSD results. The spatial structure of these modes (Figure~\ref{fig:dmd-case2-u-modes}) reveals coherent wave patterns originating from the cylinder shoulder, characteristic of vortex shedding in supersonic wakes. The V-velocity spectrum (Figure~\ref{fig:dmd-case2-v-freq}) similarly shows the shedding mode at \(St \approx 0.2\) (8~MHz) along with traces of higher frequency modes. A distinct secondary mode appears near \(St \approx 0.53\) (21~MHz), which is not a harmonic of the lower-frequency modes nor is it a artifact, since this frequency will be the dominant one in the higher Reynolds number case 3. This suggests the early development of the \(St \approx 0.5\) instability at this intermediate Reynolds number.

\begin{figure}[H]
  \centering
  \begin{subfigure}[b]{0.48\linewidth}
    \centering
    \includegraphics[width=\linewidth]{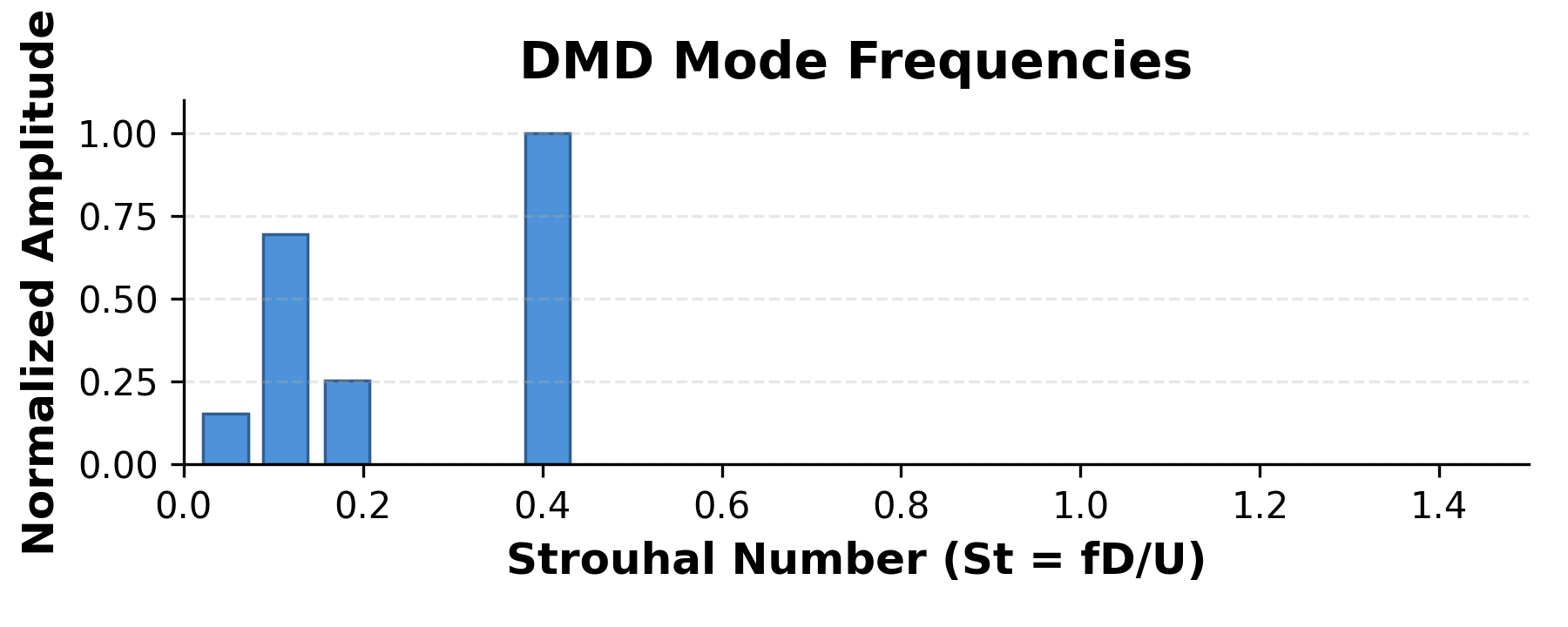}
    \caption{U-velocity: modes amplitudes}
    \label{fig:dmd-case3-u-freq}
  \end{subfigure}%
  \hfill
  \begin{subfigure}[b]{0.48\linewidth}
    \centering
    \includegraphics[width=\linewidth]{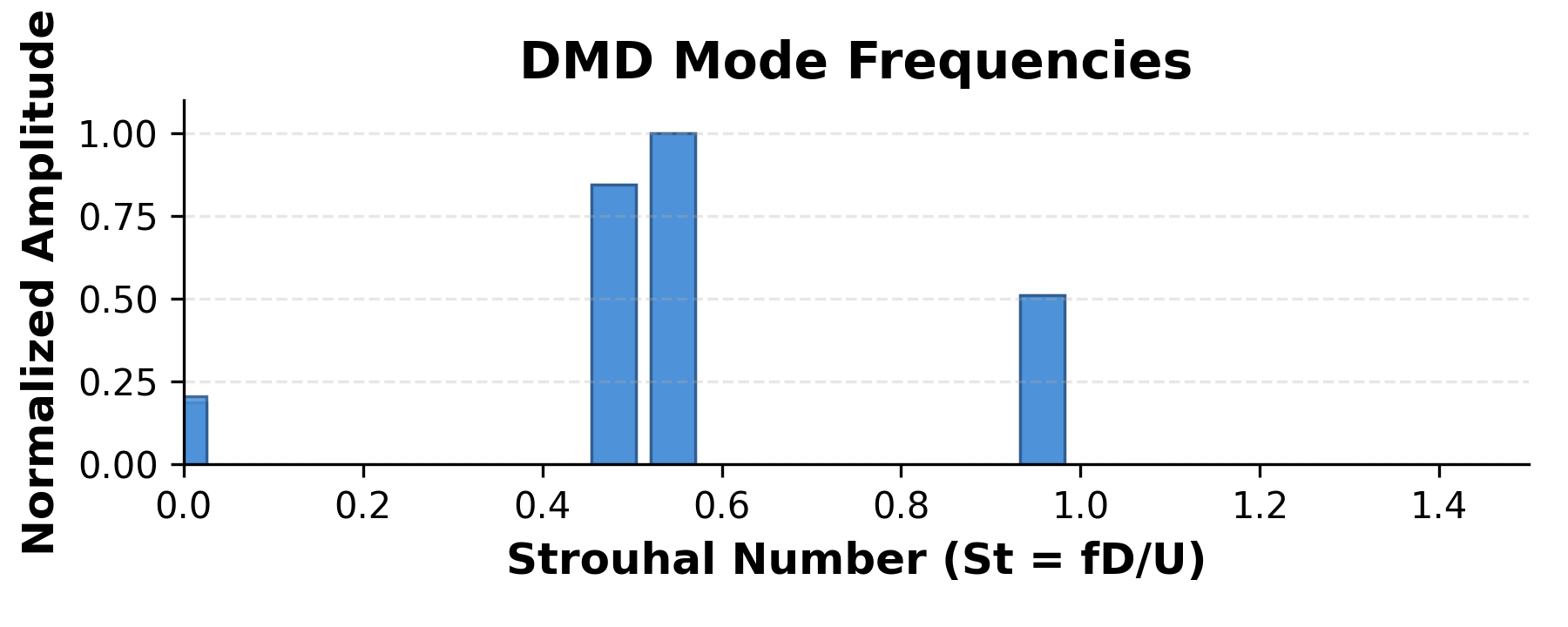}
    \caption{V-velocity: dominant mode (\(St \approx 0.55\))}
    \label{fig:dmd-case3-v-freq}
  \end{subfigure}
  \vspace{0.5em}
  \begin{subfigure}[b]{0.48\linewidth}
    \centering
    \includegraphics[width=\linewidth]{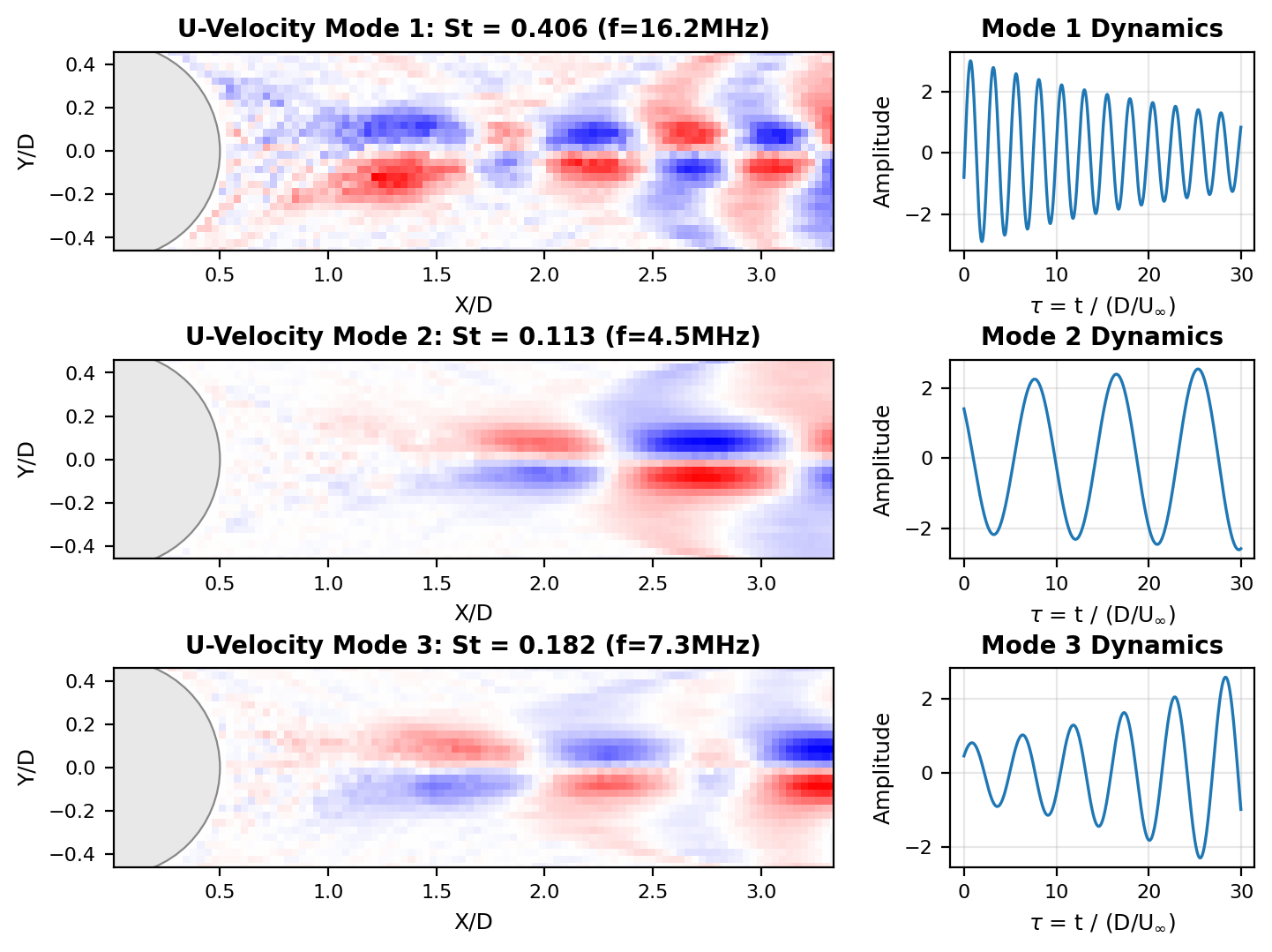}
    \caption{U-velocity modes: the high frequency mode \(St \approx 0.41\) alongside canonical vortex shedding}
    \label{fig:dmd-case3-u-modes}
  \end{subfigure}%
  \hfill
  \begin{subfigure}[b]{0.48\linewidth}
    \centering
    \includegraphics[width=\linewidth]{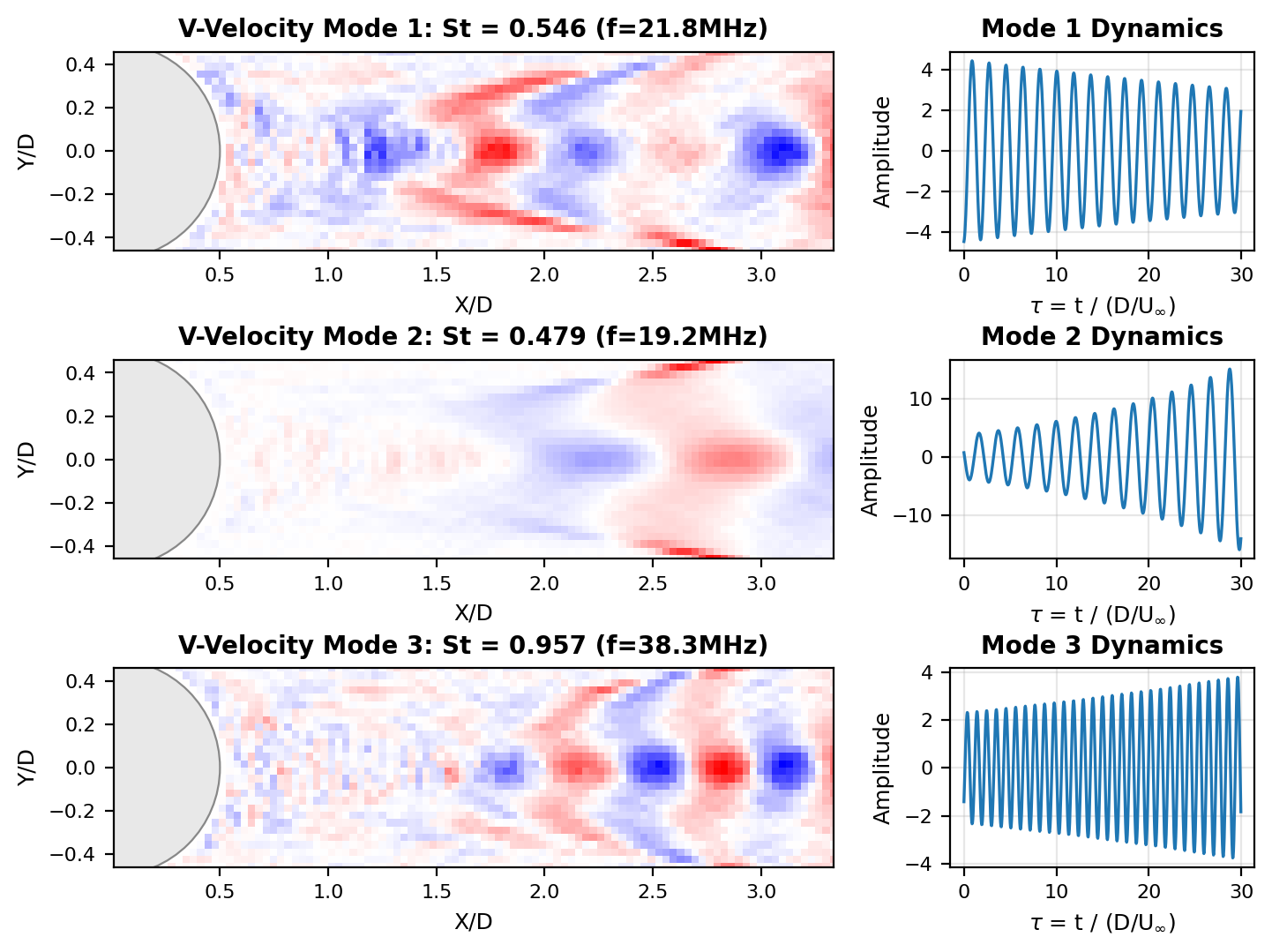}
    \caption{V-velocity modes: \(St \approx 0.5\) mode dominates, with \(St \approx 0.96\) harmonic}
    \label{fig:dmd-case3-v-modes}
  \end{subfigure}
  \caption{Wake DMD analysis for Case 3 (\(Re_D = 6750\)). Left column: streamwise (U) velocity. Right column: transverse (V) velocity. Top row: normalized mode amplitudes versus frequency. Bottom row: spatial structure and temporal dynamics of modes 1--3. The U-velocity field retains the canonical shedding mode, while the V-velocity field is now dominated by the \(St \approx 0.5\) instability.}
  \label{fig:dmd-case3-wakeNondim}
\end{figure}

Figure~\ref{fig:dmd-case3-wakeNondim} presents the corresponding results for Case 3 (\(Re_D = 6750\)). The U-velocity spectrum (Figure~\ref{fig:dmd-case3-u-freq}) again shows modes clustering around \(St \approx 0.13\)--0.18 (5--7~MHz), capturing the traces of the shedding instability, and additionally identifying the \(St \approx 0.41\) mode at 16~MHz as the strongest mode at this Reynolds number. The V-velocity results differ markedly from Case 2. The amplitude spectrum (Figure~\ref{fig:dmd-case3-v-freq}) is now dominated entirely by modes near \(St \approx 0.5\) (20~MHz). The leading V-velocity modes (Figure~\ref{fig:dmd-case3-v-modes}) oscillate at \(St \approx 0.55\) (22~MHz) and \(St \approx 0.48\) (19~MHz), capturing the same physical mode at \(St \approx 0.5\). The DMD Mode 3 at \(St \approx 0.95\) (38~MHz) is the harmonic artifact of the second DMD mode, which is apparent due to Mode 3 being exactly twice the frequency and its spatial structure being qualitatively the same as Mode 2. Thus, the V-velocity flow fluctuations are entirely dominated by the \(St \approx 0.5\) mode; this observation is consistent with the PSD results, where V-velocity fluctuations at probe 30 showed the \(St \approx 0.5\) (20~MHz) mode as the primary peak.

The comparison between Cases 2 and 3 reveals a progressive transition in wake dynamics. At \(Re_D = 1620\), both the canonical shedding mode (\(St \approx 0.2\)) and the higher-frequency \(St \approx 0.5\) mode coexist, with the former dominant in both velocity components. At \(Re_D = 6750\), the \(St \approx 0.5\) mode has grown to dominate the transverse velocity fluctuations while the shedding mode persists in the streamwise component. This suggests that at still higher Reynolds numbers, the \(St \approx 0.5\) mode would become the sole dominant instability governing wake behavior. The present results thus capture the Reynolds number range over which the wake transitions from no coherent instability (Case 1), through canonical vortex shedding (Case 2), to a regime where the \(St \approx 0.5\) mode begins to dominate (Case 3).

\section{Conclusion}
This study investigated the unsteady flow characteristics of a \SI{30}{\micro\metre} diameter cylinder in Mach 4 flow using the DSMC method, covering a Reynolds number range from 270 to 6750. The analysis identified two distinct regions of unsteadiness: the bow shock and the wake. In the stagnation region, a robust pulsation mode at 10--15~MHz was observed. Crucially, this breathing unsteadiness is found to be intrinsic to the shock and independent of the wake dynamics, persisting at a nearly constant frequency across all cases despite a 24-fold variation in number density. This suggests the mechanism is governed by the shock standoff distance rather than molecular collision timescales. In the wake, the flow transitions from a steady slip-flow regime at low Reynolds numbers to a complex unsteady state at higher Reynolds numbers (\(Re_D = 6750\)). This high-Reynolds state is characterized by the coexistence of canonical vortex shedding (\(St \approx 0.2\)) and a dominant high-frequency instability (\(St \approx 0.5\)), consistent with the aeroacoustic resonance mechanism reported in recent literature.

\section*{Acknowledgments}
The research conducted in this paper is supported by Office of Naval Research under Grant No. N000141202195 titled “Multi-scale modeling of unsteady shock-boundary layer
hypersonic flow instabilities”. This research is also supported by ACCESS Frontera supercomputer with the project number CTS23002.

\newpage
\bibliography{sample} %

\end{document}